\documentclass[11pt]{article}
\pdfoutput=1
\usepackage{jcapmod}
\usepackage{enumerate}
\usepackage{empheq}
\usepackage{feynmp}
\usepackage{booktabs}
\usepackage[english]{babel}
\usepackage{amsmath,amssymb,amsbsy,amstext, amsthm, simplewick,stackrel}
\usepackage{graphicx}
\usepackage{amsfonts}
\usepackage{amssymb}
\usepackage{caption}
\usepackage{subcaption}
\usepackage{upgreek}
 \usepackage{exscale,relsize}
 \usepackage[makeroom]{cancel}
\usepackage{float}
\usepackage{framed}
\usepackage{array}
\newcolumntype{C}[1]{>{\centering\arraybackslash}p{#1}}

\RequirePackage{color}

\usepackage{colortbl}
\definecolor{rp}{cmyk}{0.2, 1, 0.6, 0}
\definecolor{green2}{cmyk}{0, 1, 0.5, 0}
\definecolor{lightgreen}{cmyk}{0.2, 0, 0.2, 0.2}
\definecolor{lightgray}{cmyk}{0.1,0.2,0,0.1}
\definecolor{lightgray2}{cmyk}{0.4,0.4,0,0.8}
\definecolor{black}{cmyk}{1.0,1.0,1.0,1.0}

\allowdisplaybreaks[1]

\usepackage{colortbl}
\definecolor{lightgreen}{cmyk}{0.2, 0, 0.2, 0.2}
\definecolor{lightgray}{cmyk}{0.1,0.2,0,0.1}
\definecolor{lightgray2}{cmyk}{0.1,0.1,0,0.1}

\makeatletter
\newlength{\apb@width}
\newcommand{\autoparbox}[2][c]{\settowidth{\apb@width}{#2}\parbox[#1]{\apb@width}{#2}}

\makeatother

\numberwithin{equation}{section}

\def\beq{\begin{equation}}
\def\eeq{\end{equation}}

\def\bea{\begin{eqnarray}}
\def\eea{\end{eqnarray}}

\def\d{{\rm d}}

\def\beq{\begin{equation}}
\def\eeq{\end{equation}}
\def\bea{\begin{eqnarray}}
\def\eea{\end{eqnarray}}

\def\d{{\rm d}}

\def\W{{\sf W}}

\def\d{{\rm d}}

\def\H{{\cal H}}

\def\ccmb{{\chi_{{\rm \star}}}}

\def\0{{\boldsymbol 0}}
\def\k{{\boldsymbol{k}}}

\def\r{{\boldsymbol{r}}}
\def\v{{\boldsymbol{v}}}
\def\x{{\boldsymbol{x}}}

\def\n{{\hat{\boldsymbol{n}}}}

\def\fnl{f_{\mathsmaller{\rm NL}}}

\DeclareRobustCommand{\SkipTocEntry}[4]{}

\newcommand{\vev}[1]{\langle #1 \rangle}

\begin{document}

\begin{titlepage}

\setcounter{page}{1} \baselineskip=15.5pt \thispagestyle{empty}

\bigskip\

\vspace{1cm}
\begin{center}

{\fontsize{20}{28}\selectfont  \sffamily \bfseries  Efficient Evaluation of\\[10pt] Cosmological Angular Statistics}

\end{center}

\vspace{0.2cm}

\begin{center}
{\fontsize{13}{30}\selectfont Valentin Assassi, Marko Simonovi\'c and Matias Zaldarriaga}
\end{center}

\begin{center}

\vskip 8pt
\textsl{School of Natural Sciences, Institute for Advanced Study, 08540 NJ, United-States}
\vskip 7pt

\end{center}

\vspace{1.2cm}
\hrule \vspace{0.3cm}
\noindent {\sffamily \bfseries Abstract} \\[0.1cm]
Angular statistics of cosmological observables are hard to compute. The main difficulty is due to the presence of highly-oscillatory Bessel functions which need to be integrated over. In this paper, we provide a simple and fast method to compute the angular power spectrum and bispectrum of any observable. The method is based on using an FFTlog algorithm to decompose the momentum-space statistics onto a basis of power-law functions. For each power law, the integrals over Bessel functions have a simple analytical solution. This allows us to efficiently evaluate these integrals, independently of the value of the multipole~$\ell$. In particular, this method significantly speeds up the evaluation of the angular bispectrum compared to existing methods. To illustrate our algorithm, we compute the galaxy, lensing and CMB temperature angular power spectrum and bispectrum.

\vskip 10pt
\hrule

\vspace{0.6cm}
 \end{titlepage}

\tableofcontents

\newpage
%%%%%%%%%%%%%%%%%%%%
\section{Introduction}
%%%%%%%%%%%%%%%%%%%%
In cosmology, some observables such as galaxy number density or CMB anisotropies are measured on the two-dimensional sky. To compare theoretical predictions with observations, physical quantities computed in momentum space must be projected on a sphere:
\beq
\underbrace{\ \ {\cal O}_{\ell m}\ \ }_{{\rm observed}} =\ \ 4\pi i^\ell\int_0^\infty\d \chi\ W_{\cal O}(\chi) \int \frac{\d^3k}{(2\pi)^3}\ j_\ell(k\chi)\,Y_{\ell m}^*(\hat\k)\underbrace{\;{\cal O}(\k,z)\;}_{\rm computed}\ ,\label{eq:Olm0}
\eeq
where $\chi$ is the comoving distance to the observable, $z$  the corresponding redshift and $W_{\cal O}(\chi)$ is a window function.  
This projection involves an integral over a {spherical Bessel function}~$j_\ell(x)$ which, for large values of the multipole $\ell$, becomes highly oscillatory and therefore particularly difficult to integrate numerically. In some applications, it is possible to alleviate this issue by using the Limber approximation (and its improvements~\cite{LoVerde:2008re}) or the flat-sky approximation. In some other cases, this is not possible and one has to resort to numerical integration to solve the~$k$-integral~\cite{Campagne:2017xps,Lesgourgues:2011re,DiDio:2013bqa,Liguori:2010hx}.

\vskip 4pt
In this work, we provide a new efficient way of computing these integrals. Our strategy consists in projecting the statistics of ${\cal O}(\k,z)$, such as the power spectrum, onto a basis of (complex) power law functions using a simple FFTlog algorithm~\cite{Hamilton:1999uv}. For power laws in $k$, the integrals involving spherical Bessel functions can be done analytically. The full calculation then boils down to computing simple time integrals of relatively smooth functions. This significantly reduces the numerical cost of evaluating angular correlation functions. We show how to apply this method to many observables of interest such as galaxy tomography, lensing potential and CMB temperature anisotropies and, for each of them, we compute both the angular power spectrum and bispectrum. 

\vskip 10pt
\noindent{\it Note.---}All computations are done using {\sffamily Mathematica}~\cite{ram2010}. Our code can be found in the source file of the preprint of this paper. To produce the plots, we use a flat $\Lambda$CDM cosmology with $\Omega_b h^2 =0.2207 $, $\Omega_c h^2 = 0.12029 $, $h = 0.6711 $, $\tau = 0.0925$, $\Delta_\zeta^2 = 2.215\times10^{-9}$ and $n_s=0.96$. 

\section{Method and Main Results}
\label{sec:method}
\begin{figure}[h!]
\centering
                        \includegraphics[scale=0.43]{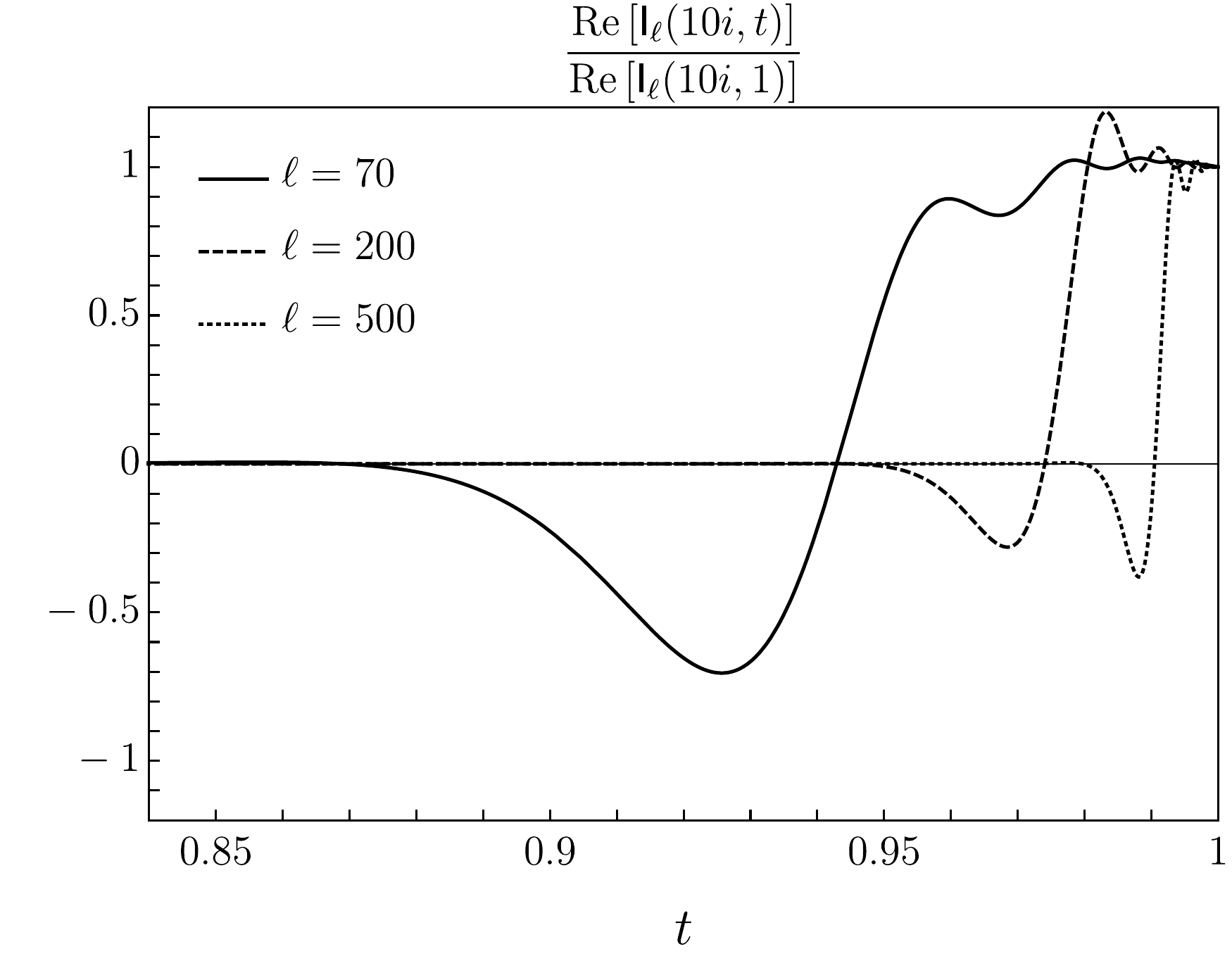} \hspace{0.2cm}  \includegraphics[scale=0.43]{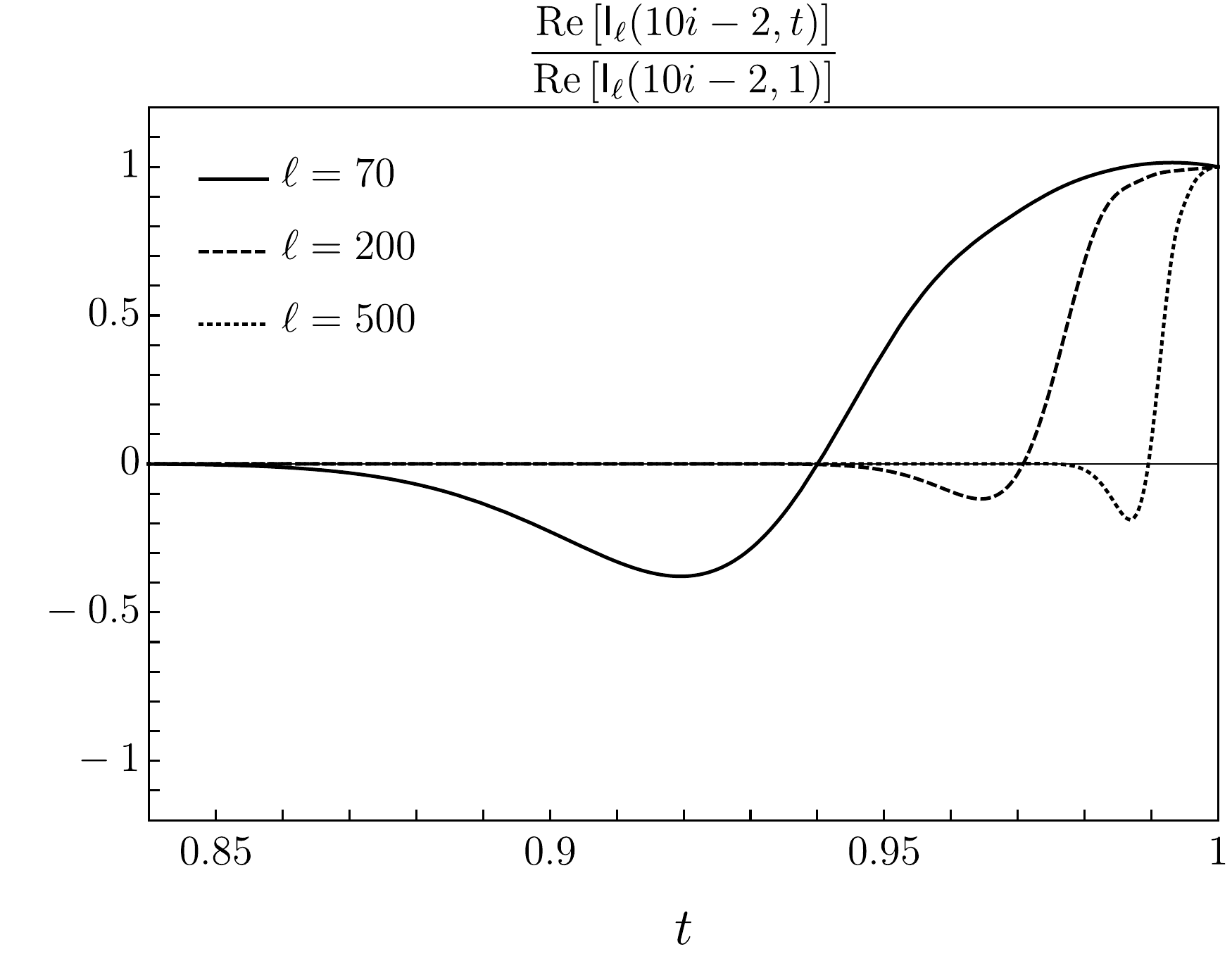}
  \caption{Plots of the real part of ${\sf I}_{\ell}(\nu,t)$ for typical values of $\ell$ and $\nu$ normalized to one at $t=1$. This function only has support around $t=1$ and becomes more sharply peaked for higher values of $\ell$. {\it Left: }The frequency $\nu=10i$ is purely imaginary. The curves have some oscillatory features. {\it Right: }The frequency $\nu=10i - 2$ has a bias of $b=2$. The curves have much less features compared to those in the left plot. }\label{fig:Iell}
 \end{figure}

In this section, we detail our method to compute the statistics of a cosmological observable projected on the sky. Let us consider an observable ${\cal O}$ (such as e.g.~galaxy number density in a given redshift bin, lensing potential or CMB fluctuations). The corresponding quantity which is observed on the sky~is
\beq
{\cal O}_{\rm obs}(\n)\ =\ \int_{0}^\infty\d \chi\ W_{\cal O}(\chi){\cal O}(\chi\n,z)\ ,\label{eq:Oobs}
\eeq
where $\chi$ is the comoving distance, $z\equiv z(\chi)$ the corresponding redshift and $W_{\cal O}(\chi)$ is a window function. In most cases, the l.h.s.~of this equation is expanded in spherical harmonics while the r.h.s.~is computed in momentum space. Using the plane wave expansion of~$e^{i\k\cdot\x}$ in spherical harmonics
\beq
e^{i\k\cdot\x}\ =\ 4\pi \sum_{\ell=0}^\infty\sum_{m=-\ell}^\ell i^\ell j_\ell(kx) Y^*_{\ell m}(\hat\k)Y_{\ell m}(\hat\x)\ ,\label{eq:pwe}
\eeq
we get the well-known projection formula:
\beq
{\cal O}_{\ell m}\  =\  4\pi i^\ell\int_0^\infty\d \chi\ W_{\cal O}(\chi) \int \frac{\d^3k}{(2\pi)^3}\ j_\ell(k\chi)\,Y_{\ell m}^*(\hat\k)\;{\cal O}(\k,z)\ .\label{eq:Olm}
\eeq
It is then straightforward to derive the expression for the $n$-point correlation function of this projected observable:
\beq
\vev{{\cal O}_{\ell_1m_1}\cdots{\cal O}_{\ell_nm_n}}\ =\ (4\pi)^n i^{\ell_{1\ldots n}}\int\left[\,\prod_{i=1}^n\d\chi_i\frac{\d^3k_i}{(2\pi)^3}\, W_{\cal O}(\chi_i)j_{\ell_i}(k_i\chi_i) Y^*_{\ell_im_i}(\hat\k_i)\right]\vev{{\cal O}_1\cdots {\cal O}_n}\ ,\label{eq:cor1}
\eeq
where $\ell_{1\ldots n}\equiv\ell_1+\cdots+\ell_n$ , ${\cal O}_i\equiv {\cal O}(\k_i,z_i)$ and the correlation functions of $\cal O$ contain an overall momentum-conserving delta function $\vev{{\cal O}_1\cdots {\cal O}_n}\equiv \vev{{\cal O}_1\cdots {\cal O}_n}'(2\pi)^3\delta_D(\k_{1}+\cdots+\k_n)$.\footnote{Even though we are considering non-equal time correlators, there is a momentum-conserving delta function as a consequence of invariance under time-independent translations.} All observables $\cal O$ can in principle be different (i.e.~eq.~(\ref{eq:cor1}) also applies for cross-correlation). However, in order to avoid clutter, we won't put any additional labels on $\cal O$ to differentiate them. 
\vskip 4pt
In this work, we only consider the power spectrum and the bispectrum and assume that both correlation functions are separable:\footnote{Our method will also work for higher-point correlation functions, provided that they can be written as products of functions of wavenumbers $k_i$ as in~(\ref{eq:sepP}) and~(\ref{eq:sepB}). Progress in that direction was made in~\cite{Regan:2010cn,Fergusson:2010ia}, where it was shown how higher-point correlation functions can be projected onto a basis of separable shapes.}
\begin{align}
\vev{{\cal O}_1{\cal O}_2}' &\,=\, f_1(k_1,z_1)f_2(k_2,z_2)\ ,\label{eq:sepP}\\
\vev{{\cal O}_1{\cal O}_2{\cal O}_3}' &\,=\, f_1(k_1,z_1)f_2(k_2,z_2)f_3(k_3,z_3) + {\rm perms}\ . \label{eq:sepB}
\end{align}
As we will see in the next two sections, these conditions are met in many cases of interest in cosmology. Usually, the power spectrum $P_{\cal O}\equiv\vev{{\cal O}_1{\cal O}_2}'$ is written as a function of only one momentum since the delta function imposes $k_1=k_2$. Similarly, the bispectrum~$B_{\cal O}\equiv\vev{{\cal O}_1{\cal O}_2{\cal O}_3}'$ is physical only when the three wavenumbers satisfy the triangle inequality. However, we will consider all wavenumbers to be independent variables and impose momentum conservation only at the very end of the calculation. One may be worried that the extension of the correlators to regions where the set of wavenumbers is not physical is not unique. For example, one could have equally chosen the r.h.s.~of~(\ref{eq:sepP}) to be $\frac{k_1}{k_2}f_1(k_1,z_1) f_2(k_2,z_2)$ since
\beq
f_1(k_1,z_1)f_2(k_2,z_2)\,\delta_D(\k_1+\k_2) = \frac{k_1}{k_2}f_1(k_1,z_1) f_2(k_2,z_2)\,\delta_D(\k_1+\k_2)\ .
\eeq
As we will see, this is however not a problem since imposing the delta function at the end will only pick up the contributions from the physical momentum regions. The delta function in the correlators is expanded in plane waves:
\beq
(2\pi)^3\delta_D(\k_{1}+\cdots+\k_n) = \int\d^3r\ e^{-i\k_{1}\cdot\r}\cdots e^{-i\k_{n}\cdot\r}\ .
\eeq
We  then use the plane wave expansion~(\ref{eq:pwe}) to write  $e^{-i\k_i\cdot\r}$ in terms of spherical harmonics and spherical Bessel functions. The angular integration over $\hat\k_i$ in~(\ref{eq:cor1}) can then be easily performed and we are left with integrals over $r$ and $k_i$. Given that the integrals in $k_i$ are separable, we get
\begin{align}
\vev{{\cal O}_{\ell_1m_1}{\cal O}_{\ell_2m_2}}&\ =\ \frac{1}{(2\pi^2)^2}\hskip 1pt\delta_{\ell_1\ell_2}\delta_{m_1m_2} \int_0^\infty\d r\ r^2\,I^{(1)}_{\ell_1}(r) I^{(2)}_{\ell_2}(r)\ , \label{eq:COP}\\
\vev{{\cal O}_{\ell_1m_1}{\cal O}_{\ell_2m_2}{\cal O}_{\ell_3m_3}}&\ =\ \frac{1}{(2\pi^2)^3}\hskip 1pt{\cal G}^{\ell_1\ell_2\ell_3}_{m_1m_2m_3} \int_0^\infty\d r\ r^2\,\big[I^{(1)}_{\ell_1}(r) I^{(2)}_{\ell_2}(r)I^{(3)}_{\ell_3}(r) + {\rm perms}\big]\ , \label{eq:COB}
\end{align}
where ${\cal G}^{\ell_1\ell_2\ell_3}_{m_1m_2m_3}$ is a geometrical factor---the Gaunt integral (see~(\ref{eq:Gaunt}))---and we have defined
\beq
{I}^{(i)}_\ell(r)\equiv  4\pi\int_0^\infty\d\chi\ W_{\cal O}(\chi)\int_0^\infty\frac{\d k}{k}\,j_{\ell}(k\chi)j_\ell(k r)\,[k^3f_i(k,z)]\ . \label{eq:CP}
\eeq
The integration in $r$ in~(\ref{eq:COP}) and~(\ref{eq:COB}) is the very operation which imposes the momentum-conserving delta function. 
The main obstacle to computing angular correlation functions boils down to evaluating the momentum integral in~(\ref{eq:CP}). The difficulty comes from the fact the Bessel function is highly oscillatory for generic values of~$r$ and $\ell$. One can find an approximate solution by noticing that the integral over $k$ peaks when $\chi=r$, since, in this case, the product of the two Bessel functions is positive and there is no cancellation in the integral. This is the well-known Limber approximation, where the integral in $k$ is replaced by a delta function:
\beq
\int_0^\infty\d k\ k^2j_\ell(k\chi)j_\ell(kr)f(k)\ \simeq\ \frac{\pi}{2r^2}f(\ell/r)\delta_D(\chi-r)\ .
\eeq
However, this approximation fails at large scales and becomes worse for narrow window functions or when cross-correlating observables whose window functions do not overlap substantially.  
\vskip 10pt
\noindent{\it Power-law decomposition.---}There is a particular example where the momentum integral in~(\ref{eq:CP}) is straightforward: whenever the function~$f_i(k,z)$ is an arbitrary power of $k$, the integral in~$k$ has a simple analytical expression which is easy to compute. This special case is particularly interesting since the decomposition of any function in terms of power laws can be naturally achieved by performing a simple Fourier transform of $k^3f_i(k,z)$ in $\log k$. Numerically, this can be done using the well-known FFTlog algorithm~\cite{Hamilton:1999uv}. This is the central idea of this work, which draws inspiration from the recently developed algorithm to efficiently evaluate loops in large-scale structures~\cite{Schmittfull:2016jsw,McEwen:2016fjn}.  
 \vskip 4pt
As we will see in Sections~\ref{sec:2pt} and~\ref{sec:3pt}, the function~$f_i(k,z)$ in~(\ref{eq:CP}) is either an integer power of~$k^2$ or contains a transfer function which is suppressed for large values of $k$. First, we focus on the latter case 
 (the former case is considered at the very end of this section)  where the integrand in~(\ref{eq:CP}) has mainly support in a finite momentum range~$[k_{\rm min},k_{\rm max}]$ (with $k_{\rm min}>0$). In this range, $k^3 f_i(k,z)$ can be decomposed as follows: 
\beq
k^3 f_i(k,z)= \sum_{n=-\infty}^\infty c^{(i)}_{n}(z)\ k^{\nu_n}\quad{\rm where}\quad \nu_n\equiv\frac{2\pi i}{\Delta\kappa}n-b\quad {\rm and} \quad \Delta\kappa\equiv\log(k_{\rm max}/k_{\rm min})\ .\label{eq:FFTlog}
\eeq
The coefficients of this decomposition are
\beq
c^{(i)}_n(z) \equiv \frac{1}{\Delta\kappa}\int_{\kappa_{\rm min}}^{\kappa_{\rm max}}\d\kappa\ e^{(3+b)\kappa}f_i(e^\kappa,z)\,e^{-\frac{2\pi i n}{\Delta\kappa} \kappa}\ ,\label{eq:coef}
\eeq
where $\kappa\equiv \log k$. Note that in~(\ref{eq:FFTlog}) and (\ref{eq:coef}) we have introduced a real number $b$ which we refer to as the {\it ``bias''}~\cite{Schmittfull:2016jsw,McEwen:2016fjn}. This extra piece comes about when we Fourier transform~$k^{3+b}f_i(k,z)$ instead of~$k^{3}f_i(k,z)$ . While this bias can in principle take any value, it needs to be chosen with care: The Fourier decomposition in~(\ref{eq:FFTlog}) matches the true function~$k^{3+b}f_i(k,z)$ only in the range $[k_{\rm min},k_{\rm max}]$. Outside this range, this is no longer true (since the r.h.s.~of (\ref{eq:FFTlog}) is periodic in $\log k$). Hence, we need to ensure that the kernel multiplying $k^{3+b}f_i(k,z)$ in~(\ref{eq:CP}) suppresses both IR and UV contributions. This is indeed the case when $-2<b<2\ell$, since the kernel has the following behavior in the IR and the UV:\footnote{The spherical Bessel functions have the following properties: 
\begin{align}
&\lim_{x\to0}\, j_{\ell}(x)\;\propto\; x^\ell\qquad{\rm and}\qquad \lim_{x\to\infty}\, j_{\ell}(x)\;\propto\; x^{-1} \ .
\end{align}}
\begin{align}
&\lim_{k\to0}\ [k^{-(1+b)}j_\ell(k\chi)j_\ell(kr)]\ \propto\ k^{2\ell-1-b}\ ,\label{eq:BesselIR}\\
&\lim_{k\to\infty}[k^{-(1+b)}j_\ell(k\chi)j_\ell(kr)]\ \propto\ k^{-(3+b)}\ .\label{eq:BesselUV}
\end{align}  
Finally, let us make three comments about this decomposition: $(i)$ First, notice that the powers~$\nu_n$ are {\it complex} numbers. Hence, this formula can easily capture features in the transfer functions such as e.g.~BAO wiggles. $(ii)$ Only a few of the lowest frequencies $\nu_n$ contribute a significant amount to the sum (\ref{eq:FFTlog}).  The number of frequencies which need to be kept will depend on the types of features in $f_i(k,z)$, the value of the bias $b$ and the precision we want to achieve. In practice, this number is at most~$100$. $(iii)$ In order to make the function $f_i(k,z)$ smooth at the boundary of the interval $[k_{\rm min},k_{\rm max}]$, it is usual to modify $f_i(k,z)$ by apodization or zero-padding~(see e.g.~\cite{McEwen:2016fjn}). However, in practice one can apply the FFTlog without any preprocessing. This is because the boundary effects are strongly suppressed due to the rapid decay of the spherical Bessel functions both in the IR and the UV (see eqs.~(\ref{eq:BesselIR}) and (\ref{eq:BesselUV})).
\vskip 10pt
\noindent{\it Power-law solution.---}We now explain how the decomposition~(\ref{eq:FFTlog}) can be used to evaluate~(\ref{eq:CP}). 
 Substituting (\ref{eq:FFTlog}) into (\ref{eq:CP}), we can write $I_{\ell}^{(i)}(r)$ as a simple sum:\footnote{Notice that the dependence of $c_n$ on the redshift $z$ has been relabelled by a dependence on the comoving distance along the line of sight $\chi$.}
\beq
I^{(i)}_{\ell}(r) = \sum_{n}\int_0^\infty\d\chi\ W_{\cal O}(\chi)\,c^{(i)}_{n}(\chi)\;{\chi}^{-\nu_n}\,{\sf I}_{\ell}\big(\nu_n,\tfrac{r}{\chi}\big)\ ,\label{eq:CP2}
\eeq
where
\beq
{\sf I}_{\ell}(\nu,t)\, \equiv\, 4\pi\int_0^\infty\d v\ v^{\nu-1}j_{\ell}(v)j_{\ell}(vt)\ ,\label{eq:Cellnu}
\eeq
and we have defined $v\equiv k\chi$ and $t\equiv r/\chi$. Remarkably, despite the oscillatory nature of the Bessel function, ${\sf I}_{\ell}(\nu,t)$ is a smooth function which can be calculated analytically:\footnote{Some analytical results involving integrals of Bessel functions have also been studied in~\cite{Fergusson:2006pr, delaBella:2017qjy}.}
\beq
{\sf I}_\ell(\nu,t)\ =\  \frac{2^{\nu-1}\pi^2\,\Gamma(\ell+\tfrac{\nu}{2})}{\Gamma(\tfrac{3-\nu}{2})\Gamma(\ell+\tfrac{3}{2})}\,
t^\ell\;_2F_1\left(\tfrac{\nu-1}{2},\ell+\tfrac{\nu}{2},\ell+\tfrac{3}{2},t^2\right) \quad {\rm for\ } t\leq1\ , \label{eq:Iellnu2}
\eeq
where ${}_2F_1(\cdots\hskip-0.5pt)$ is the hypergeometric function (whose precise definition can be found in~Appendix~\ref{app:hyper}). Using the definition and properties of the hypergeometric function, the function~${\sf I}_\ell(\nu, t)$ can be efficiently evaluated (see~Appendix~\ref{app:Iell}). More precisely, in {\sffamily Mathematica} (using compiled functions) the function ${\sf I}_\ell(\nu, t)$ can be computed within approximately $10^{-4}\,{\rm s}$ (which is comparable to the time required to evaluate any elementary function such as e.g.~sine or exponential). Furthermore, this function satisfies a recursion relation allowing to evaluate ${\sf I}_{\ell}(\nu,t)$ in terms of lower-order multipoles (see Appendix~\ref{app:Iell}), which can additionally speed up the evaluation. To compute ${\sf I}_{\ell}(\nu,t)$ in the region $t>1$, one can simply use the following property:
\beq
{\sf I}_\ell\big(\nu,\tfrac{1}{t}\big) \ = \ t^\nu\hskip 1pt {\sf I}_\ell(\nu,t) \ . \label{eq:Iprop}
\eeq
The function~${\sf I}_\ell(\nu,t)$ is plotted as a function of $t$ for different values of $\ell$ and $\nu$ in fig.~\ref{fig:Iell}. Let us make two comments about these plots. First, we notice that the rapid oscillations of the Bessel functions have disappeared. There are some residual features which can be reduced by appropriately choosing the value of the bias (the higher the bias, the less feature). Second, for relatively large multipoles  (more precisely for $\ell \gtrsim {\cal O}(10) $), the function  ${\sf I}_\ell(\nu,t)$ has mainly support around $t=1$ and the peak sharpens with increasing values of $\ell$. In this regime, we recover the Limber approximation where the r.h.s.~of~(\ref{eq:Cellnu}) becomes proportional to $\delta_D(t-1)$.
\vskip 10pt
\noindent{\it Increasing the bias.---}Given that when the frequency $\nu$ has a large negative real part, the amount of features in~${\sf I}_\ell(\nu,t)$ is reduced (see fig.~\ref{fig:Iell}), one should Fourier transform $k^3f_i(k,z)$ with the largest possible bias. This, however, usually comes at the cost of increasing the number of frequencies in order to reach the same precision. Fortunately, there is another way of changing the real part of the frequency. Indeed, one can increase the powers of $k^2$ by noticing that the spherical Bessel function satisfies
\beq
\left[\frac{\partial^2}{\partial \chi^2}+\frac{2}{\chi}\frac{\partial}{\partial \chi}-\frac{\ell(\ell+1)}{\chi^2}\right]j_\ell(k\chi) = -k^2j_\ell(k\chi)\ .\label{eq:Helm}
\eeq
We can then apply the differential operator which appears on the l.h.s.~of this equation to the Bessel function in~(\ref{eq:CP2}) and integrate by part the derivatives. It is then straightforward to show that
\begin{align}
I^{(i)}_\ell(r)&\ = \ \sum_{n} \left[{\rm BT}(\nu_n)+\int_0^\infty\d\chi\ {\cal D}_\ell [W_{\cal O}(\chi)\,c^{(i)}_{n}(\chi)]\;{\chi}^{-(\nu_n-2)}\,{\sf I}_{\ell}(\nu_n-2,\tfrac{r}{\chi})\right], \label{eq:Ishift}
\end{align}
where ${\rm BT}(\nu)$ are boundary terms. In general, the window functions and their derivatives vanish at the boundary so that ${\rm BT}(\nu) = 0$. However, in some applications, this is not the case and one has to keep track of these boundary terms (see \S\ref{sec:lensingbisp}). In~(\ref{eq:Ishift}), we introduced the differential operator ${\cal D}_\ell$ defined as
\beq
{\cal D}_\ell \equiv -\frac{\partial^2}{\partial \chi^2}+\frac{2}{\chi}\frac{\partial}{\partial \chi}+\frac{\ell(\ell+1)-2}{\chi^2}\ .\label{eq:Dell}
\eeq
Notice that in~this equation, all frequencies are effectively shifted by $-2$, keeping the coefficients~$c_n^{(i)}$ {\it unchanged}. This is therefore different from adding a bias $b$ in~(\ref{eq:FFTlog}), since in this case the coefficients depend on the value of the real part of the frequency. Finally let us point out that one can keep applying the same differential operator to obtain as large a (negative) real part of the frequency~$\nu_n$ as required. However, when the window function doesn't have an analytical expression, computing the successive derivatives can be challenging as any small error gets magnified by the differential operator (see \S\ref{sec:CMBbisp}). 
\vskip 10pt
\noindent{\it Dirac delta function.---}Finally, let us comment on another important result which will prove particularly useful when computing the galaxy and lensing bispectrum. So far, we have examined the case where~$f_i(k,z)$ contains a transfer function. However, in some examples (e.g.~for the galaxy and lensing bispectrum) this function can be an integer power of $k^2$ so that the integral we need to solve is 
\beq
\int_0^\infty\d k\ k^{2n}j_{\ell}(k\chi)j_{\ell}(k r)\ ,
\eeq
where $n$ is an integer. For $n\leq 0$, the integral is convergent\footnote{To be more precise, the integral is convergent only for $n > -\ell$, which is always the case in practice.  } and one can apply (\ref{eq:Iellnu2}). On the other hand, when $n>0$, this integral is divergent. For $n=1$, there is a well-known solution to this integral in terms of a delta function~\cite{Arfken:2005:MMP}:
\beq
\int_0^\infty\d k\ k^{2}j_{\ell}(k\chi)j_{\ell}(k r) = \frac{\pi}{2r^2}\delta_D(\chi-r)\ .
\eeq
Using~(\ref{eq:Helm}), it is then easy to show that for $n\geq 1$, we have
\beq
\int_0^\infty\d k\ k^{2n}j_{\ell}(k\chi)j_{\ell}(k r) = \frac{\pi}{2r^2}\left[-\frac{\partial^2}{\partial \chi^2}-\frac{2}{\chi}\frac{\partial}{\partial \chi}+\frac{\ell(\ell+1)}{\chi^2}\right]^{n-1}\delta_D(\chi-r)\ .\label{eq:delta}
\eeq
This result is particularly useful since, in this case, the function $I_{\ell}(r)$ is no longer an integral but just the (derivatives of the) window function evaluated at $\chi=r$.
\vskip 10pt
\noindent{\it Summary.---}In this section, we presented all the building blocks needed to evaluate angular power spectra and bispectra. We showed that computing~(\ref{eq:CP}) comes down to performing an FFTlog and a time integration of a smooth function. The number of operations to evaluate the integrals over Bessel functions is the same as the number of frequencies, which is about 100. To estimate how much improvement our method yields with respect to a direct integration, this should be compared with the number of sampling points in $k$ needed in other integration schemes. We will give more details in the examples considered next. In Sections~\ref{sec:2pt} and \ref{sec:3pt}, we apply the method developed here to several cosmological observables, giving detailed explanations about the implementation of this algorithm for each example. All the results computed in the following sections are done using a {\sffamily Mathematica} code which can be found in the source file of this paper.

\section{Angular Power Spectrum}
\label{sec:2pt}
In this section, we study the angular power spectrum:
\begin{align}
\vev{{\cal O}_{\ell_1m_1}{\cal O}_{\ell_2m_2}}&\, =\, (4\pi)^2 i^{\ell_{1}+\ell_2}\int\left[\,\prod_{i=1}^2\d\chi_i\frac{\d^3k_i}{(2\pi)^3}\, W_{\cal O}(\chi_i)j_{\ell_i}(k_i\chi_i) Y^*_{\ell_im_i}(\hat\k_i)\right]\nonumber\\
&\hskip 200pt \times \vev{{\cal O}_1{\cal O}_2}'\,(2\pi)^3\delta_D(\k_1+\k_2)\ .\label{eq:Cl1}
\end{align}
Assuming that the power spectrum satisfies~(\ref{eq:sepP}) and defining $\vev{{\cal O}_{\ell_1m_1}{\cal O}_{\ell_2m_2}}\equiv \delta_{\ell_1\ell_2} \delta_{m_1m_2} C_\ell^{\mathsmaller{({\cal O})}}$, eq.~(\ref{eq:COP}) then simply becomes
\beq
C_\ell^{\mathsmaller{({\cal O})}}\ =\ \frac{1}{(2\pi^2)^2} \int_0^\infty\d r\ r^2\,[I^{\mathsmaller{({\cal O})}}_{\ell}(r)]^2\ ,\label{eq:Cell0}
\eeq
where $I^{\mathsmaller{({\cal O})}}_\ell(r)$ was defined in~(\ref{eq:CP}). One can then proceed as explained in Section~\ref{sec:method}. 
\vskip 4pt
However, there is a slightly different  (and more common)  way of evaluating the two-point function which can sometimes  be faster and easier to implement.  Starting with~(\ref{eq:Cl1}), we first integrate over~$\k_2$ and then over $\hat \k_1$ to get
\beq
C_\ell^{\mathsmaller{({\cal O})}}\, =\, \frac{2}{\pi} \int_0^\infty\d\chi\int_0^\infty \d\chi'\ W_{\cal O}(\chi)W_{\cal O}(\chi')\int_0^\infty\frac{\d k}{k}\ j_{\ell}(k\chi)j_{\ell}(k\chi')\,[k^3 P_{\cal O}(k,z,z')]\ . \label{eq:Cell1}
\eeq 
As before, we expand $k^3 P_{\cal O}(k,z,z')$ in power laws using~(\ref{eq:FFTlog}) and write the integral in $k$ in terms of the function~${\sf I}_{\ell}(\nu,\chi'/\chi)$. After a change of variables, $\chi'\equiv\chi t$, we get\footnote{Notice that, as opposed to (\ref{eq:FFTlog}), the coefficients $c_n$ of the Fourier transform of $k^3 P_{\cal O}(k,z,z')$ now depends on both reshifts $z$ and $z'$ (or equivalently, on the comoving distances $\chi$ and $\chi'$).}
\beq 
C_\ell^{\mathsmaller{({\cal O})}}\, = \, \frac1{2\pi^2} \sum_{n} \int_0^\infty \d\chi \int_0^\infty \d t \ c_n(\chi,\chi t)\chi^{1-\nu_n} W_{\cal O}(\chi)W_{\cal O}(\chi t) {\sf I}_{\ell} (\nu_n,t) \;.
\eeq
In practice, integrating $t$ from 0 to $\infty$ may require too many sampling points, particularly at low $\ell$ where ${\sf I}_\ell(\nu,t)$ has a very broad support and can oscillate. Numerically, it is much more efficient to map the region $t>1$ to $0<t<1$, using the relation~(\ref{eq:Iprop}). We can then write\footnote{For auto-correlation, the first and second terms in the second line of this equation yield the same result. In this case, we only need to compute the integral with only one of them and multiply the whole result by two.}
\begin{align}
C_\ell^{\mathsmaller{({\cal O})}}&\, = \, \frac1{2\pi^2} \sum_{n}  \int_0^1 \d t\ {\sf I}_{\ell} (\nu_n,t) \nonumber\\
&\hskip 60pt\times \int_0^\infty \d\chi \ \chi^{1-\nu_n}W_{\cal O}(\chi)\left[ c_n(\chi,\chi t) W_{\cal O}(\chi t) +t^{\nu_n-2}c_n(\chi,\chi/t)W_{\cal O}(\chi/t)\right] \ .\label{eq:Cell2}
\end{align}
The only difference between~(\ref{eq:Cell0}) and (\ref{eq:Cell2}) is that, in the former case, we have one integral along the line of sight $\chi$ and one integral in $r$, while in the latter there are two integrals along the line of sight. These two approaches, however, usually require a similar number of operations. The most efficient method depends on the specificities of the problem in hand. For example, the computational advantage of~(\ref{eq:Cell2})  over (\ref{eq:Cell0}) comes from the fact that, in the former case, the function ${\sf I}_\ell(\nu,t)$ needs to be evaluated fewer times. However~(\ref{eq:Cell2}) requires more FFTlog (one for each value of $\chi$ and $t$) unless the power spectrum  is separable in space and time. In all the examples of this section, we will use (\ref{eq:Cell2}) and reserve the approach with integration over $r$ for the bispectrum in Section~\ref{sec:3pt}.
\vskip 4pt
Finally, let us point out that the difference between the exact result and the Limber approximation is more pronounced on large angular scales where cosmic variance is high. Therefore, one may wonder whether this difference is statistically relevant. This was recently studied in the context of weak lensing~\cite{Lemos:2017arq,Kitching:2016zkn,Kilbinger:2017lvu}. Moreover, in some cases (such as in the presence of local primordial non-Gaussianity) the difference between the two is enhanced on large scales and therefore using the Limber approximation could bias the parameters inferred from data. For these reasons, we would like to quantify how distinct the two approaches are by evaluating the signal-to-noise ratio~(SNR):
\beq
({\rm SNR}_{P})^2\ \equiv\ \sum_{\ell=\ell_{\rm min}}^{\ell_{\rm max}}\frac{\big(C_{\ell,\rm exact}^{ \mathsmaller{({\cal O})}}-C_{\ell,\rm limber}^{  \mathsmaller{({\cal O})}}\big)^2}{(\Delta C_{\ell}^\mathsmaller{({\cal O})})^2}\ ,\label{eq:SNR}
\eeq
where $\Delta C_\ell^\mathsmaller{({\cal O})}\equiv \sqrt{\tfrac{2}{2\ell+1}}\,C_\ell^\mathsmaller{({\cal O})}$ is the full-sky cosmic variance. In this section, we compute the SNR for the galaxy and lensing power spectra. Notice that we do not include any noise in the SNR as we are only interested in the fundamental limit imposed by cosmic variance.

\subsection{Galaxy Tomography}
\label{sec:PSg}

\begin{figure}[h!]
\centering
                       \includegraphics[scale=0.43]{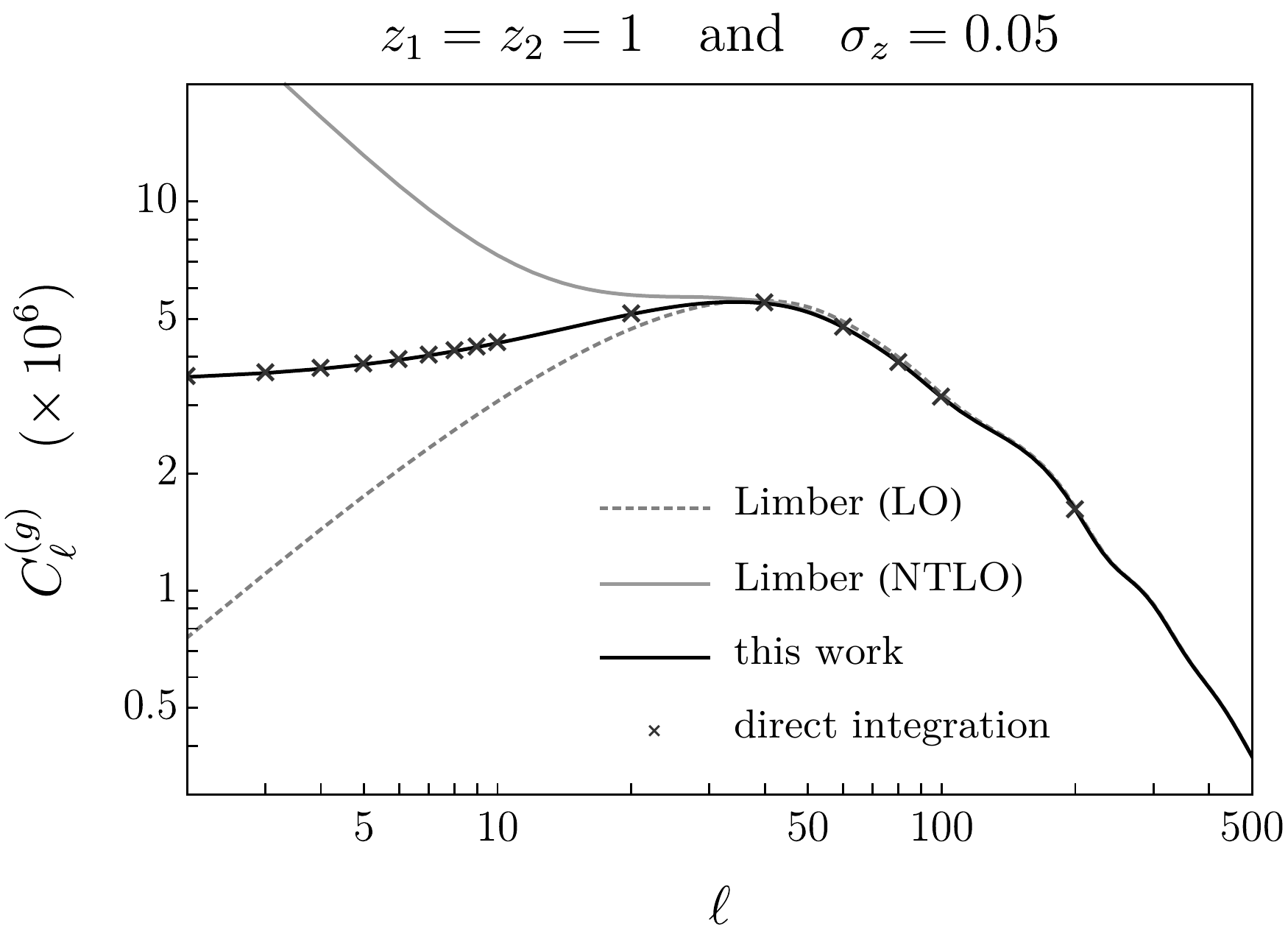} \hspace{0.2cm}  \includegraphics[scale=0.43]{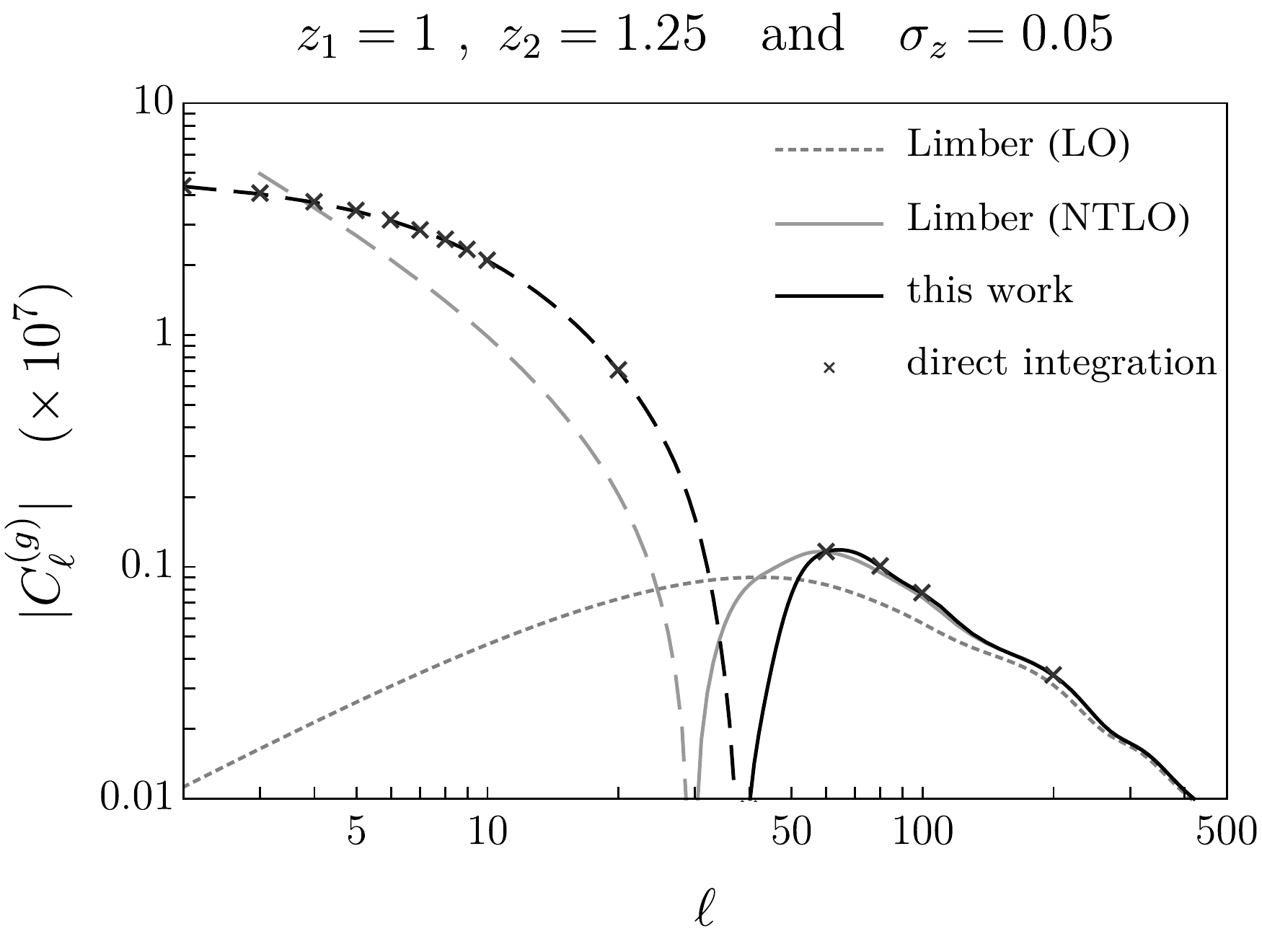}\\[20pt]
                       \includegraphics[scale=0.43]{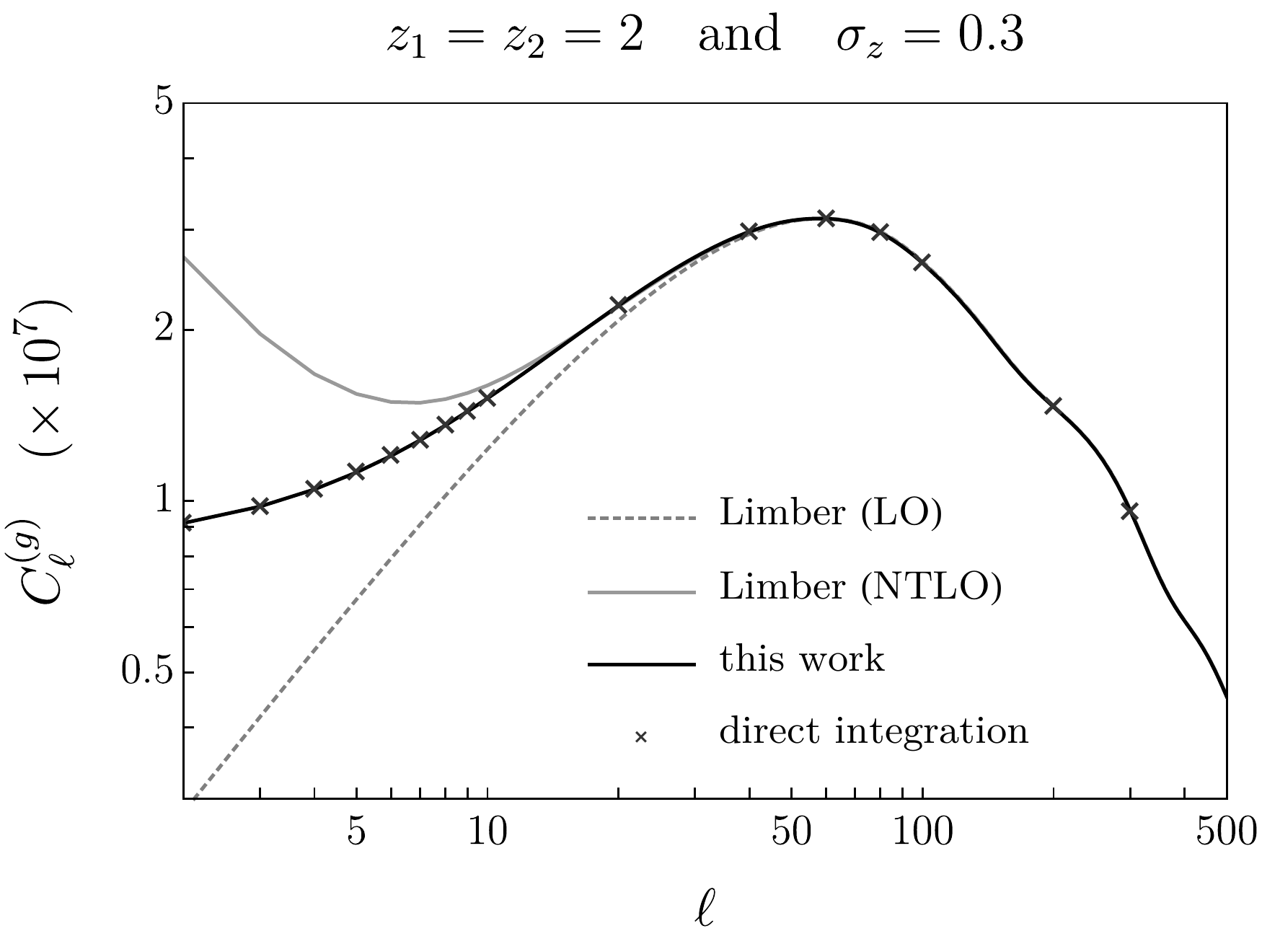} \hspace{0.2cm}  \includegraphics[scale=0.43]{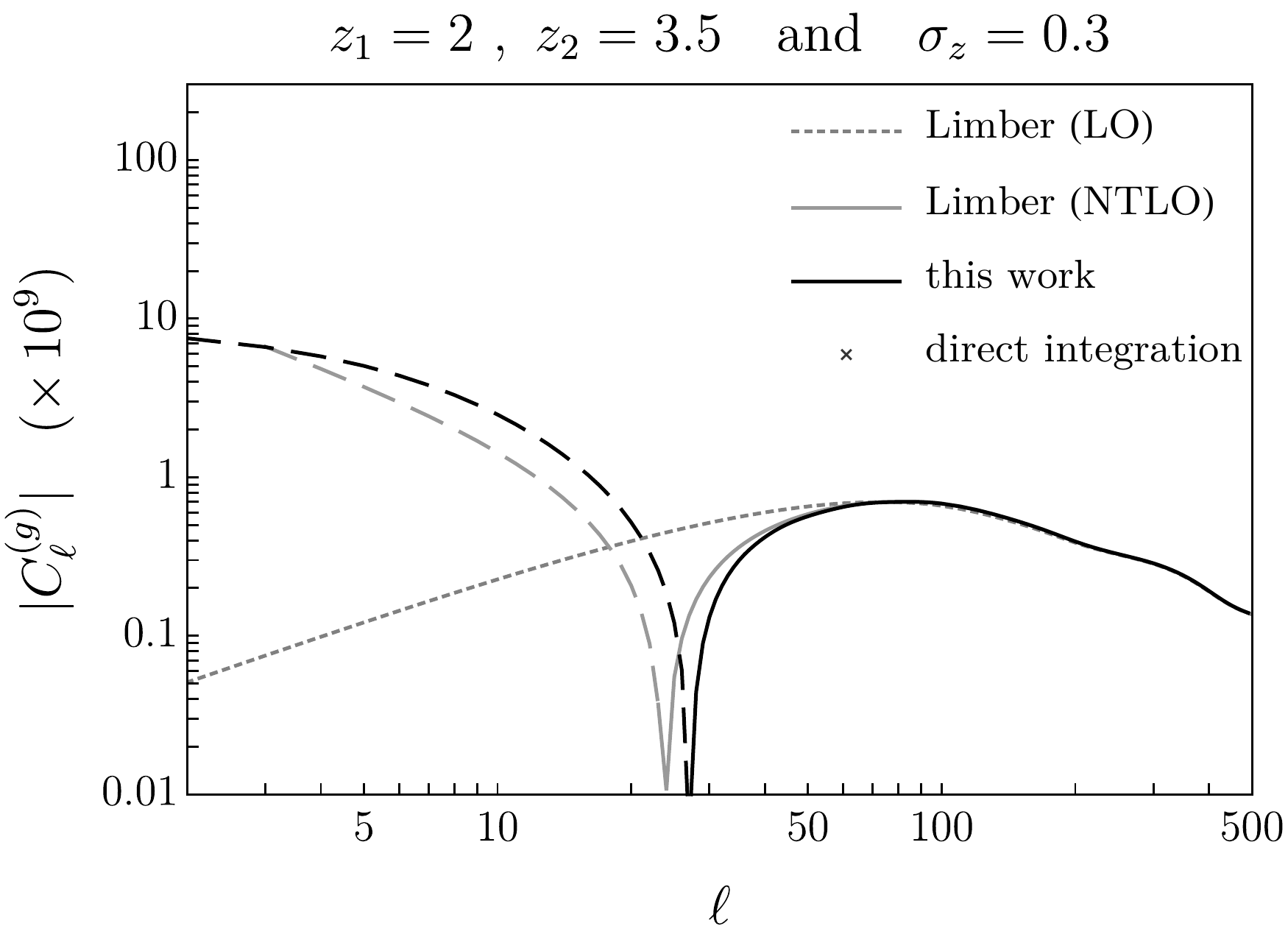}
  \caption{Galaxy angular power spectrum with $b_1=1$ and a window function of the form~(\ref{eq:Wg}), using the leading order (LO) and next-to-leading order (NLO) Limber approximation (gray small-dashed curve and gray solid curve, respectively),  our method (black solid curve), and direct integration (black crosses). The gray and black long-dashed curves correspond to negative values. The upper panel corresponds to a thin redshift bin with $\sigma_z=0.05$. The top left plot represents the power spectrum evaluated at~$z_1=z_2=1$ while the top right plot represents the cross-correlation of two redshift bins $z_1=1$ and $z_2=1.25$. The lower panel corresponds to a wider redshift bin with $\sigma_z=0.3$. The bottom left plot represents the power spectrum evaluated at $z_1=z_2=2$ while the bottom right plot represents the cross-correlation of two redshift bins~$z_1=2$ and~$z_2=3.5$. To produce these plots, we used the following parameters:  $N_\nu = 100$ frequencies in the FFTlog with a bias of $b = 1.9$. For the line-of-sight integrals, we used $N_\chi = N_t=50$ sampling points.}
 \label{fig:Cg}
 \end{figure}

We first apply our method to the galaxy angular power spectrum. In particular, it is clear how~(\ref{eq:Cell2}) can be directly applied to the full nonlinear galaxy power spectrum $P_{g}(k,z,z')$ (obtained from simulations or perturbative calculations). 
However, nonlinear corrections only contribute at small scales where the Limber and flat-sky approximations give fairly accurate results. Therefore we will focus on large scales where linear theory applies. In this regime, the galaxy overdensity~$\delta_g(\x,z)$ is related to the matter overdensity by a local linear bias:
\beq
\delta_g(\x,z) = b_1\,\delta(\x,z)\ ,\label{eq:linbias}
\eeq
where $b_1$ is the linear clustering bias and $\delta(\x,z)$ is the matter overdensity. For simplicity, we assumed that the bias is redshift-independent, but our approach also works for time-dependent bias.  In the linear regime, the matter overdensity is simply~$\delta(\x,z) = D(z)\delta_{in}(\x)$ where~$D(z)$ is the linear growth factor and $\delta_{in}(\x)$ the initial fluctuations whose statistics is given by the initial power spectrum~$P_{in}(k)\equiv\vev{\delta_{in}(\k)\delta_{in}(-\k)}'$.
\vskip 4pt
Galaxies are in general measured in a given redshift bin characterized by a window function which depends on the survey. For concreteness, we use a window function which is Gaussian along the line of sight: 
\beq
W_g(\chi,\bar\chi,\sigma_\chi)\, \equiv\,\frac{1}{\sqrt{2\pi}\sigma_\chi}\exp\left[-\frac{(\chi-\bar\chi)^2}{2\sigma_\chi^2}\right]\ .\label{eq:Wg}
\eeq
This window function corresponds roughly to a Gaussian redshift bin with center $\bar z\equiv z(\bar\chi)$ and width $\sigma_z\simeq \sigma_\chi/H(\bar z)$. We then calculate the galaxy power spectrum~$C_\ell^{(g)}$ using~(\ref{eq:Cell1}): 
\begin{align}
C_\ell^{(g)}&\ =\ \frac{2}{\pi} b_1^2\int_0^\infty\d\chi\int_0^\infty\d\chi'\ \W_g^{(1)}(\chi)\W_g^{(1)}(\chi')\int_0^\infty\frac{\d k}{k}j_\ell(k\chi)j_\ell(k\chi')\,[k^3P_{in}(k)]\nonumber\\[5pt]
&\ =\ \frac{2}{\pi} b_1^2\int_0^\infty\d\chi\int_0^\infty\d\chi'\ [{\cal D}_\ell\W_g^{(1)}(\chi)][{\cal D}_\ell\W_g^{(1)}(\chi')]\int_0^\infty\frac{\d k}{k}j_\ell(k\chi)j_\ell(k\chi')\,[k^{-1}P_{in}(k)]\ ,\label{eq:Cellg}
\end{align}
where $\W_g^{(n)}(\chi)\equiv W_g(\chi,\bar\chi,\sigma_\chi)[D(\chi)]^n$ and we have used~(\ref{eq:Helm}) to decrease the powers of $k$ in the integrand. This last step is motivated by the fact that not only $k^{-1}P_{in}(k)$ is easier to Fourier transform than $k^{3}P_{in}(k)$ but also it allows us to work with a large bias, which improves the convergence of the integrals along the line of sight (see Section~\ref{sec:method}). 
\vskip 10pt
\noindent{\it Performance.---}The fact that our method allows to solve the $k$-integral in (\ref{eq:Cellg}) {\it analytically}, should lead to a significant improvement over numerical integration. To compare the efficiency of our approach with direct integration or other numerical methods, we calculated the galaxy power spectrum with Dirac-delta window functions:
\beq
C_\ell^{(g)}(z_1,z_2)\ =\ \frac{2}{\pi} b_1^2D(\chi_1)D(\chi_2)\int_0^\infty k^2\d k\ j_\ell(k\chi_1)j_\ell(k\chi_2)P_{in}(k)\ .
\eeq
Using $N_\nu=100$ frequencies in the FFTlog, we evaluated every multipole up to $\ell_{\rm max} = 1000$ within $0.2\,{s}$ on a single core using our {\sf Mathematica} code. We have checked that this time is independent of the values of~$z_1$ and~$z_2$. Furthermore, using a recursion relation which relates the functions ${\sf I}_\ell(\nu,t)$ with different values of $\ell$ (see.~(\ref{eq:recursionIl})), we were able to reduce the computing time by a factor of 10. Finally, let us emphasize that these performances are the same for both auto and cross-correlations. This is significantly faster than advanced numerical integration algorithms such as {\sffamily Angpow} (which performs the same computation in $0.38\,{\rm s}$ for auto-correlation and~$0.76\,{\rm s}$ for cross-correlation (see Test 1 and Test 2 of Table~1 in~\cite{Campagne:2017xps})).
\vskip 20pt
%\va{move this sentence to comments?} 
\noindent{\it Results.---}Our results are plotted in fig.~\ref{fig:Cg}, where our method is compared with the Limber approximation and direct numerical integration for different redshift bins. Let us make a few comments:
\begin{itemize}
%\item Our method allows to solve the $k$-integral {\it analytically}. To compare the efficiency of this approach with direct integration or other numerical methods, we calculated the galaxy power spectrum with Dirac-delta window functions:
%\beq
%C_\ell^{(g)}(z_1,z_2)\ =\ \frac{2}{\pi} b_1^2D(\chi_1)D(\chi_2)\int_0^\infty k^2\d k\ j_\ell(k\chi_1)j_\ell(k\chi_2)P_{in}(k)\ .
%\eeq
%Using $N_\nu=100$ frequencies, we evaluated every multipole up to $\ell_{\rm max} = 1000$ within $0.2\,{s}$ on a single core using our {\sf Mathematica} code, independently of the value of the redshifts~$z_1$ and~$z_2$. Furthermore, using a recursion relation which relates the functions ${\sf I}_\ell(\nu,t)$ with different values of $\ell$ (see~(\ref{eq:recursionIl})), we were able to reduce the computing time by a factor of 10. Finally, let us emphasize that these performances are the same for both auto and cross-correlations. This is significantly faster than advanced numerical integration algorithms such as {\sffamily Angpow} (which performs the same computation in $0.38\,{\rm s}$ for auto-correlation and~$0.76\,{\rm s}$ for cross-correlation (see Test 1 and Test 2 of Table~1 in~\cite{Campagne:2017xps})).
\item In order to make these plots, we used the following parameters: $N_\nu = 100$ frequencies in the log-Fourier transform~(\ref{eq:FFTlog}) and $N_\chi = N_t =50$ sampling points for the integrals in $\chi$ and~$t$. The integration along the line of sight is computed using a Gauss quadrature which scales as~${\cal O}(N_\chi\times N_t)$. We chose these parameters as $(i)$ increasing these numbers does not change the final result and $(ii)$~our method agrees with direct numerical integration to within $0.01\%$. Importantly (and unlike previous methods) these numbers do not change for cross-correlations. Concretely, the galaxy angular power spectrum can be computed for about 200 multipoles within 30 seconds on a laptop using our {\sffamily Mathematica} code on a single core. We expect that this time strongly depends on implementation: a more dedicated code, written in better-suited programing languages (such as {\sffamily C} or {\sffamily Fortran}), which uses more optimized quadrature for the line-of-sight integrals and implements recursion relations (that relates the functions~${\sf I}_\ell(\nu,t)$ with different multipoles---see Appendix~\ref{app:Iell}) can further reduce the overall computing time. 
\item Notice that the Limber approximation becomes better with wider redshift bins. This is expected: the Limber approximation assumes that the integral in $k$ in~(\ref{eq:Cellg}) can be replaced by a delta function~$\delta_D(\chi-\chi')$. This is due to the fact that the integral over~$k$ yields a function which (for high $\ell$) sharply peaks at $\chi=\chi'$. However, this approximation is justified provided that the window function $W_{\cal O}$ is wider than the width of this peak. 
\item For cross-correlation of different redshift bins, the Limber approximation predicts almost zero signal. This comes from the fact that, by definition, the Limber approximation only takes into account contributions from the correlation of observables evaluated at the same redshift. Of course, in reality there are contributions from long-wavelength modes across redshift bins which produce some signal at low $\ell$. 
\item One could also consider next-to-leading order (NLO) corrections to the Limber approximation~\cite{LoVerde:2008re}:
\beq
C_\ell^{(g)} = b_1^2\int_0^\infty\label{eq:beyondLim}\frac{\d\chi}{\chi} P_{in}\left(\nu/\chi\right)\left[[\hat\W_g(\chi)]^2-\frac{\hat\W_g(\chi)}{\nu^2}\left(\chi^2\hat\W_g''(\chi)+\frac{\chi^3}{3}\hat\W_g'''(\chi)\right) \right] \ ,
\eeq
where $\hat\W_g(\chi)\equiv \W_g^{(1)}(\chi)/\sqrt{\chi}$ and $\nu\equiv \ell+1/2$. 
 While these corrections improve the result at high multipoles, they break down at low multipoles. 
 % Notice that the next-to-leading order term (NLO) in (\ref{eq:beyondLim}) can become large at small $\ell$. 
 For window functions that peak around~$\bar\chi$ and have a width of~$\sigma_\chi$, the typical multipole at which the NLO becomes comparable to the leading-order (LO) term is $\ell_{\mathsmaller{\rm NLO}}\,\sim\, (\bar \chi/\sigma_\chi )^{3/2}$. Therefore, for $\ell \lesssim  \ell_{\mathsmaller{\rm NLO}}$, higher-order corrections do not improve the Limber approximation, as illustrated in fig.~\ref{fig:Cg}. In general, we expect improvements  at all multipoles from higher-order corrections only for very broad window functions (i.e.~when~$\sigma_\chi \sim \bar\chi$).
  %More precisely, for a window function with support around $\bar \chi$ and width $\sigma_\chi \ll \bar\chi$, next-order corrections break at $\ell\simeq (\bar \chi/\sigma_\chi )^{3/2}$.
\item Finally let us comment on the signal-to-noise ratio~(\ref{eq:SNR}). Its value depends on several parameters: the width of the window functions and whether we consider auto or cross-correlation. We computed the SNR for auto-correlation for thin ($\sigma_z=0.05$ and $z=1$) and wide ($\sigma_z=0.3$ and $z=2$) window functions. When summing over the first 500 multipoles, we found that ${\rm SNR}_P\simeq3$ and ${\rm SNR}_P\simeq5$ for the thin and wide window function, respectively. This shows that using the Limber approximation may affect the values of cosmological parameters inferred from galaxy surveys which cover a big fraction of the sky by a few standard deviations. For cross-correlations, the SNR is much higher, showing that the Limber result is a much worse approximation than for auto-correlation. These computations of the SNR assume a full sky survey so that $\ell_{\rm min}=2$. Of course, for current surveys, which only cover a fraction of the sky, $\ell_{\rm min}$ is larger and the SNR smaller.
\end{itemize}
\subsubsection*{Redshift-Space Distortions}
\begin{figure}[h!]
\centering
                       \includegraphics[scale=0.43]{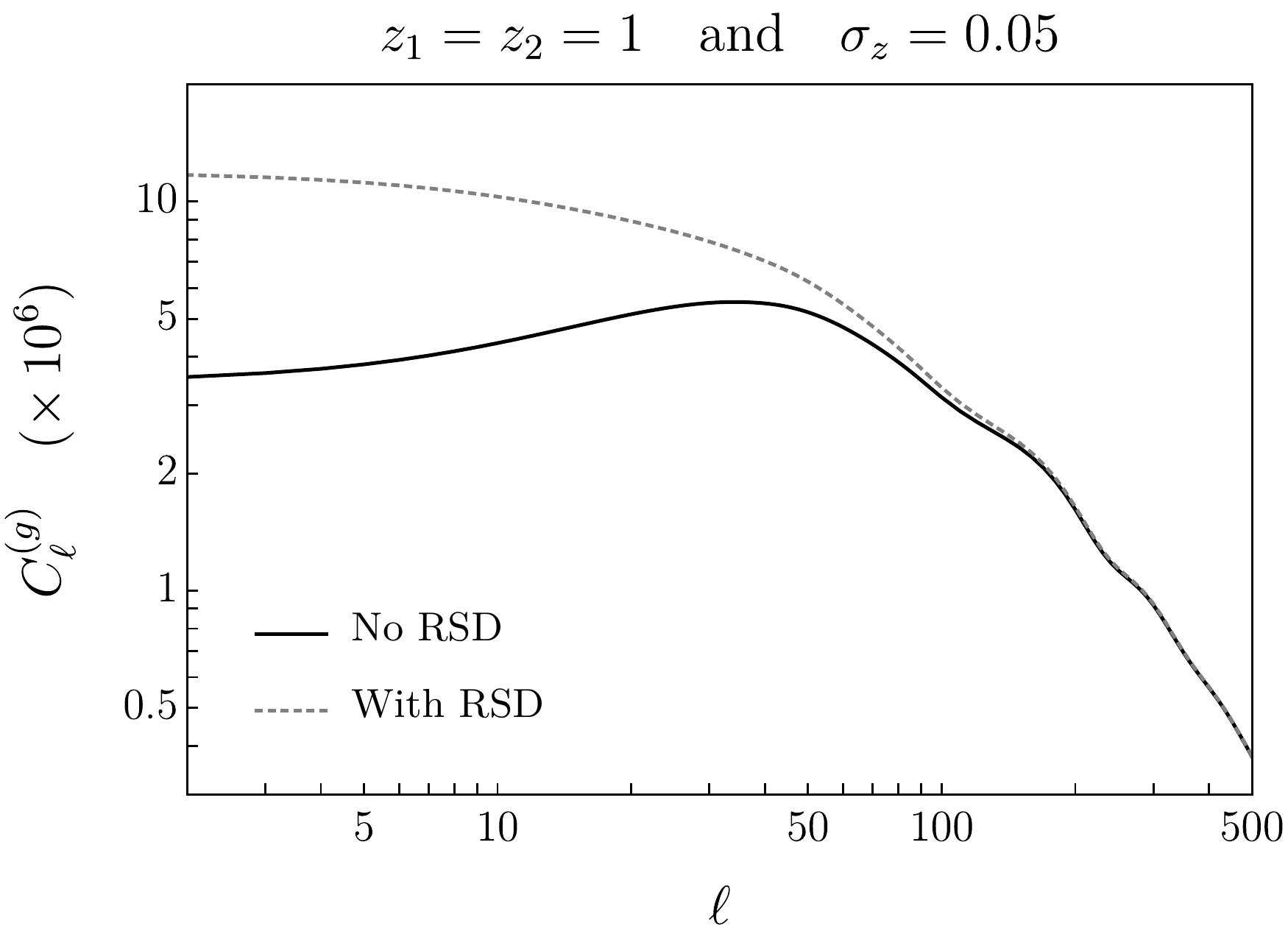}
                         \caption{Galaxy angular power spectrum assuming $b_1=1$ and a window function of the form~(\ref{eq:Wg}) with (gray dashed curve) and without (solid black curve) RSD. To produce this plot, we used the following parameters:  $N_\nu = 100$ frequencies in the FFTlog with a bias of $b = 1.9$. For the line-of-sight integrals, we used $N_\chi = N_t=50$ sampling points.}\label{fig:RSD}
 \end{figure}

The peculiar velocity of galaxies distorts the galaxy density which is measured in surveys, such that the observed galaxy overdensity is
\beq
\delta_g^{(\rm obs)}(\n) = \int_0^\infty\d\chi\ W_g(\chi,\bar\chi,\sigma_\chi)\left[\delta_g-\frac{1}{\H}\partial_\chi (\n\cdot\v)\right](\chi\n,z) \ ,%\left[\delta_g-\frac{1}{\H}\frac{\partial (\n\cdot\v)}{\partial\chi}\right]\ ,
\eeq
where $\H$ is the comoving Hubble parameter and $\v$ is the comoving velocity of galaxies.  At leading order in perturbation theory, the velocity of galaxies matches that of dark matter which, at linear order, is~(e.g.~\cite{Kaiser:1987qv})
\beq
\v(\k,z) = i\frac{\k}{k^2}f(z)\H(z) D(z)\delta_{in}(\k)\ ,
\eeq
where $f\equiv \d\ln D/\d\ln a$. We can then show that the spherical harmonics component of the observed galaxy overdensity is (see e.g.~\cite{Bonvin:2014xia})
\beq
[\delta_g^{\rm (obs)}]_{\ell m} = 4\pi i^\ell\int_0^\infty\d\chi\ \W^{(1)}_g(\chi)\int\frac{\d^3k}{(2\pi)^3}\left[b_1 j_\ell(k\chi) - f(\chi)j''_\ell(k\chi)\right] Y_{\ell m}^*(\hat\k)\delta_{in}(\k)\ .\label{eq:dgRSD}
\eeq
A direct way to compute the correlation of $[\delta_g^{\rm (obs)}]_{\ell m}$ is to use the fact that $j_\ell''$ is a linear combination of $j_\ell$ and $j_{\ell+1}$. Then, the method developed in Section~\ref{sec:method} still applies provided that the integral (\ref{eq:Cellnu}) is generalized to include the convolution of two Bessel functions with different multipoles $\ell$:
\begin{align}
{\sf I}_{\ell_1,\ell_2}(\nu,t)&\ \equiv\ 4\pi\int_0^\infty\d v\ v^{\nu-1}j_{\ell_1}(v)j_{\ell_2}(vt)\ ,\label{eq:I2ell}
%\\
%&\ =\ \frac{2^{\nu -1}\pi ^2\,  \Gamma
%   (\frac{\ell_1+\ell_2+\nu}{2}) \,
%   }{\Gamma
%(\frac{3-\nu+\ell_1-\ell_2}{2})\Gamma
% (\ell_2+\frac{3}{2}) }\,t^{\ell_2} \;_2F_1\left(\tfrac{\nu -1-\ell_1+\ell_2}{2},\tfrac{\ell_1+\ell_2+\nu }{2},\ell_2+\tfrac{3}{2},t^2\right)  \quad {\rm for\ } t\leq1\ .
\end{align}
which can also be expressed in terms of a hypergeometric function\footnote{More precisely, this integral has the following solution:
\beq
{\sf I}_{\ell_1,\ell_2}(\nu,t)\ =\ \frac{2^{\nu -1}\pi ^2\,  \Gamma
   (\frac{\ell_1+\ell_2+\nu}{2}) \,
   }{\Gamma
(\frac{3-\nu+\ell_1-\ell_2}{2})\Gamma
 (\ell_2+\frac{3}{2}) }\,t^{\ell_2} \;_2F_1\left(\tfrac{\nu -1-\ell_1+\ell_2}{2},\tfrac{\ell_1+\ell_2+\nu }{2},\ell_2+\tfrac{3}{2},t^2\right)  \quad {\rm for\ } t\leq1\ .
\eeq
}. 
An alternative approach is to slightly massage eq.~(\ref{eq:dgRSD}): using (\ref{eq:Helm}) and integrating by part several times with respect to the line-of-sight variable $\chi$:
\beq
[\delta_g^{\rm (obs)}]_{\ell m} = 4\pi i^\ell\int_0^\infty\d\chi\ \left[b_1{\cal D}_\ell - \frac{\d^2}{\d\chi^2}f(\chi)\right]\W^{(1)}_g(\chi)\int\frac{\d^3k}{(2\pi)^3} \frac{1}{k^2}j_\ell(k\chi)Y_{\ell m}^*(\hat\k)\delta_{in}(\k)\ .
\eeq
Hence, the effect of redshift-space distortions (RSD) on large scales is captured by simply adding an extra term in the window function. Usually, integrating by part is simpler and more efficient (as it requires the same number of operations as computing the galaxy angular power spectrum without RSD) but in some applications (e.g.~for very narrow window functions) it may be more convenient to use~(\ref{eq:I2ell}). In fig.~\ref{fig:RSD}, we've plotted the galaxy angular power spectrum with and without RSD. 

\subsubsection*{Primordial non-Gaussianity}
\begin{figure}[h!]
\centering
                       \includegraphics[scale=0.43]{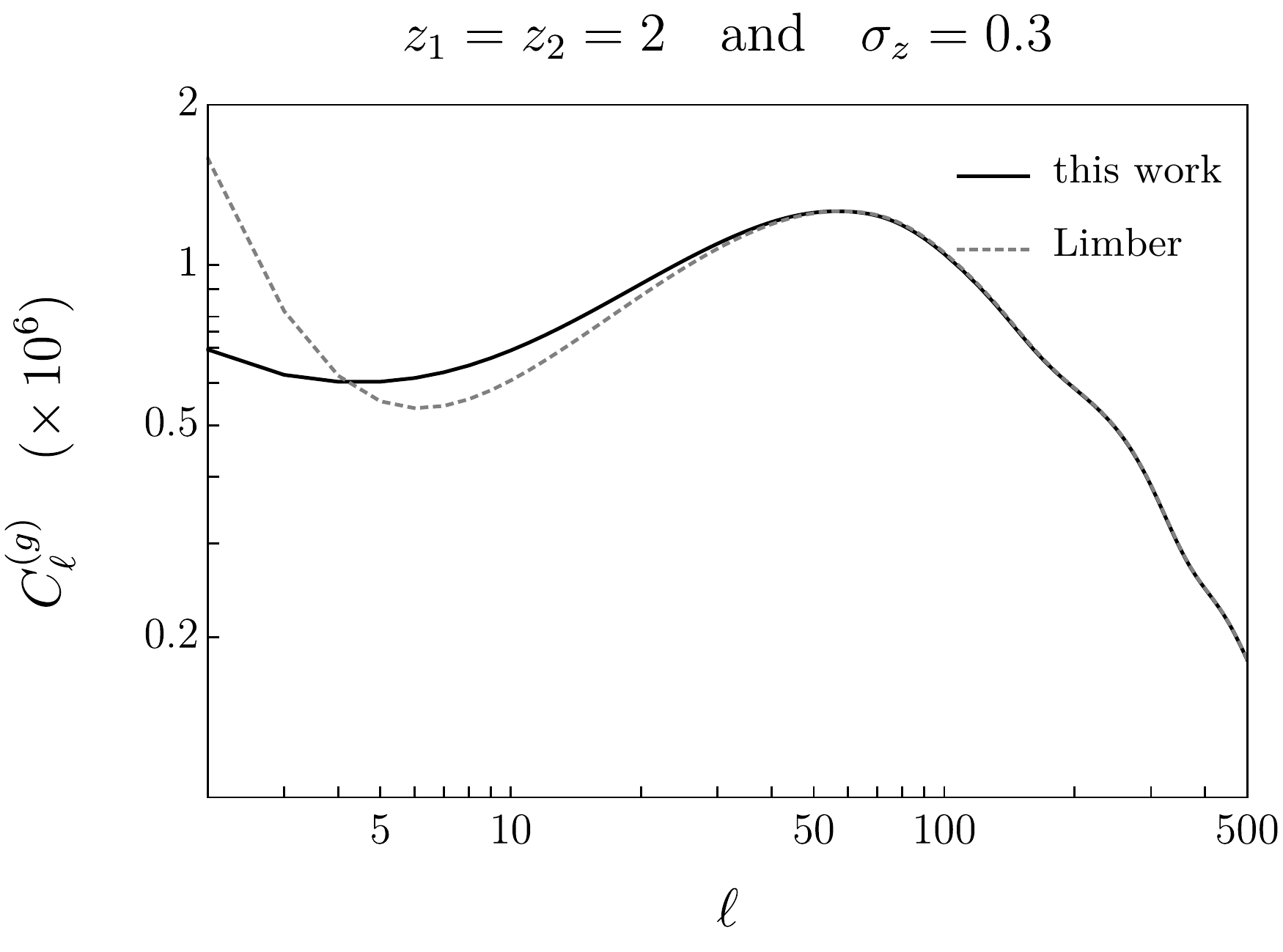} \hspace{0.2cm}  \includegraphics[scale=0.43]{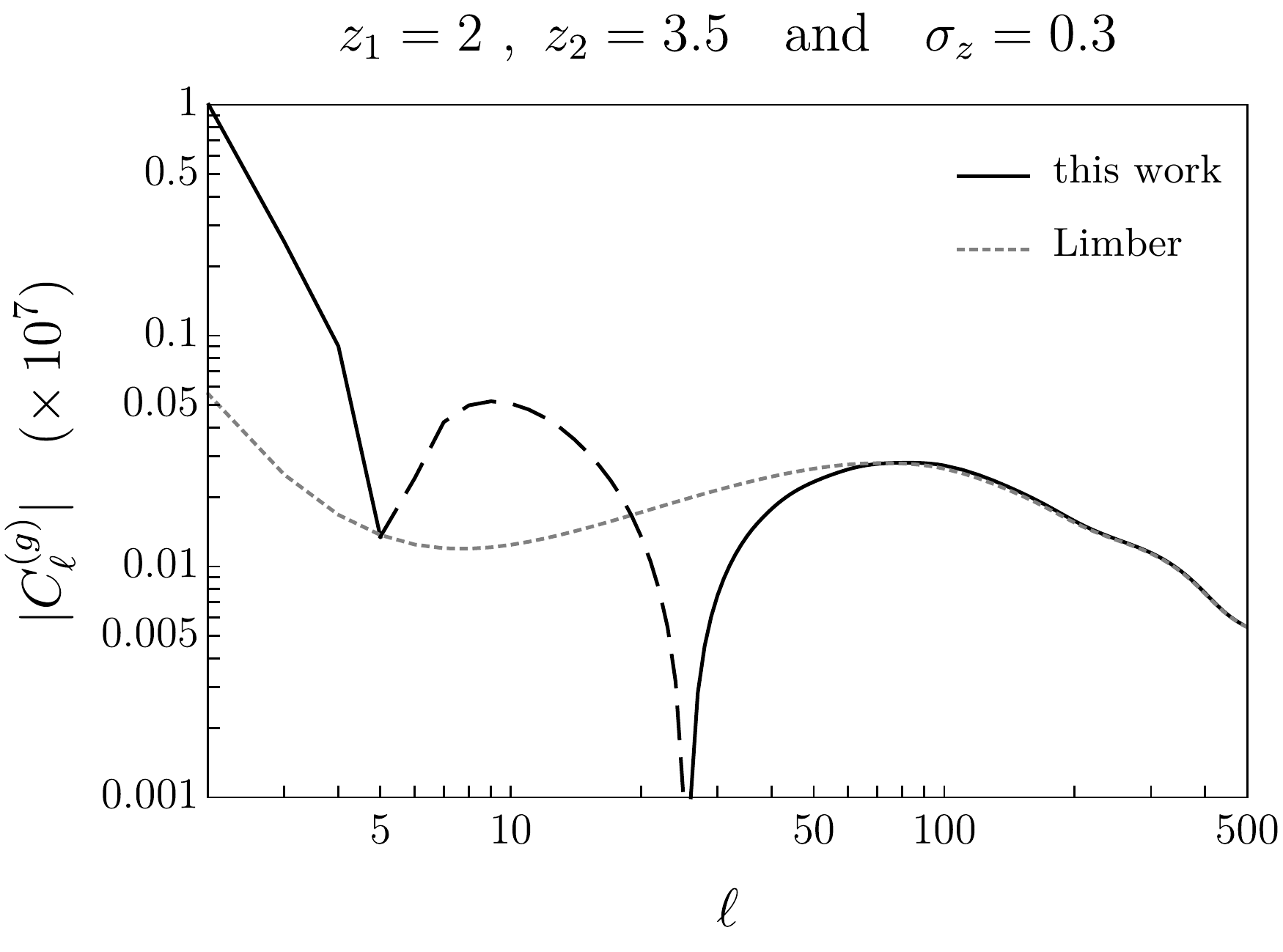}
  \caption{Galaxy angular power spectrum with scale-dependent bias induced by local PNG. We considered a survey with $\sigma_z = 0.3$ and have assumed a constant bias with value $b_1 = 2$. The amplitude of PNG is $\fnl =10$. {\it Left:} power spectrum at $z_1=z_2=2$. {\it Right:} cross-correlation of redshift bins  $z_1=2$ and $z_2=3.5$. The black dashed curve corresponds to negative values. To produce these plots, we used the following parameters:  $N_\nu = 100$ frequencies in the FFTlog with a bias of $b = 1.9$. For the line-of-sight integrals, we used $N_\chi = N_t=50$ sampling points.}\label{fig:CgSD}
 \end{figure}

Finally, let us conclude this section with a particularly useful application of our method: the scale-dependent bias induced by local primordial non-Gaussianity (PNG). In~\cite{Dalal:2007cu}, it was shown  that in the presence of PNG, the large-scale galaxy power spectrum is
\beq
P_{g}(k,z) = \big[b_1+\fnl \Delta b(k,z)\big]^2\,[D(z)]^2P_{in}(k)\ ,
\eeq
where $\Delta b(k,z)$ is the scale-dependent bias
\beq
\Delta b(k,z)\,\simeq\, 2\delta_c(b_1-1)\frac{\Omega_mH_0^2}{k^2}(1+z)\ ,
\eeq
where $\delta_c\simeq 1.6$ is the critical density of spherical collapse, $\Omega_m$ the matter density parameter and~$H_0$ the Hubble parameter. 
This scale-dependent bias contributes at large angular scales where the Limber approximation fails as illustrated in fig.~\ref{fig:CgSD}: for auto-correlation power spectra, the Limber approximation overestimates the effect of the scale-dependent bias, while for cross-correlation it significantly underestimates it. This difference may be important as  cross-correlations of galaxies with different redshifts can contain a significant fraction of the signal for local PNG.  

\subsection{Gravitational Lensing}
\label{sec:PSlensing}

\begin{figure}[h!]
\centering
\includegraphics[scale=0.55]{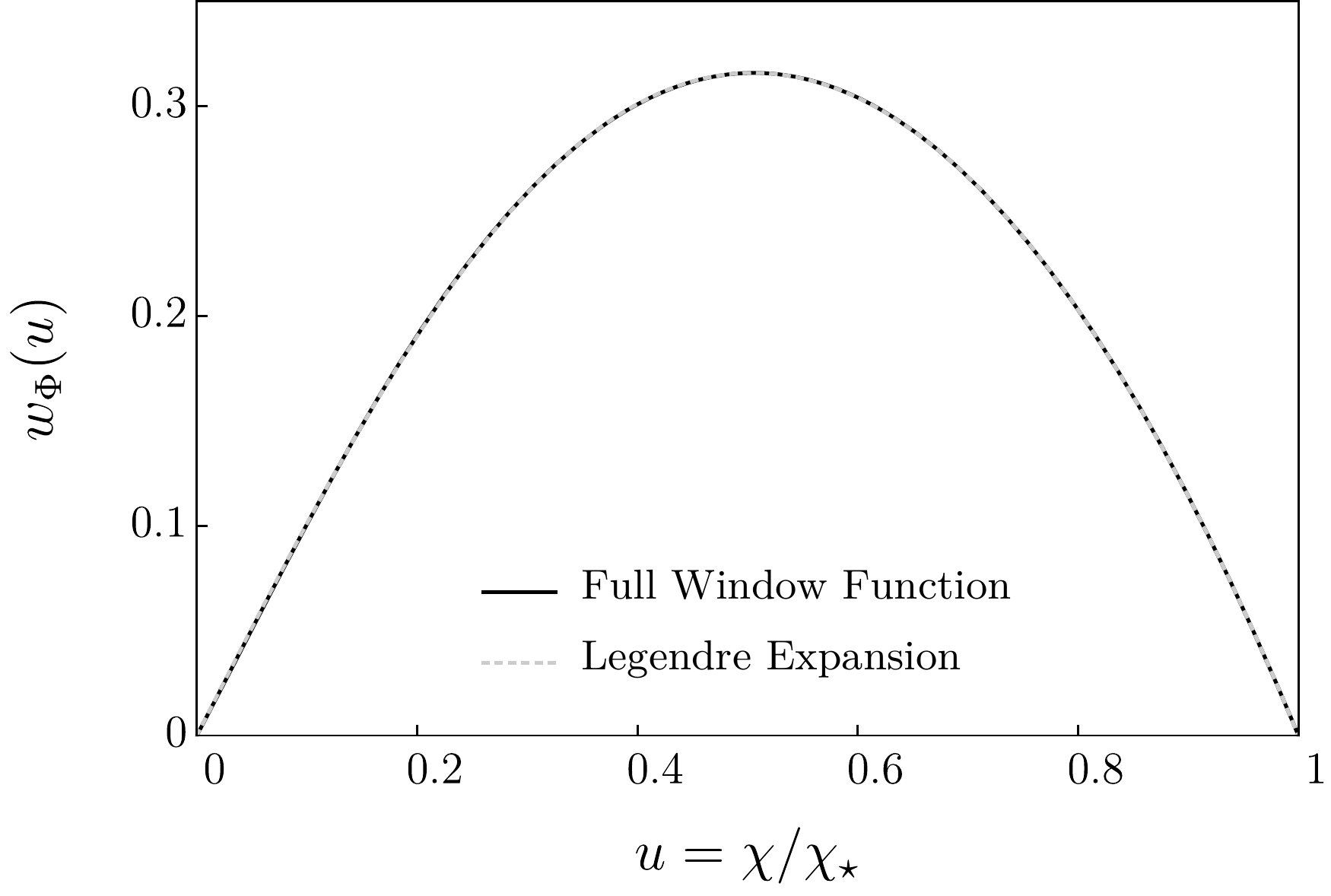}
\caption{Rescaled lensing window function $w_\Phi(u)$ (see~(\ref{eq:window})). The solid black curve represents the actual window function (l.h.s.~of~(\ref{eq:Legendre})) while the dotted gray curve corresponds to the Legendre expansion (r.h.s.~of ~(\ref{eq:Legendre}) with $p_{\rm max}=31$). \label{fig:window}}
\end{figure}

\begin{figure}[h!]
\centering
\includegraphics[scale=0.43]{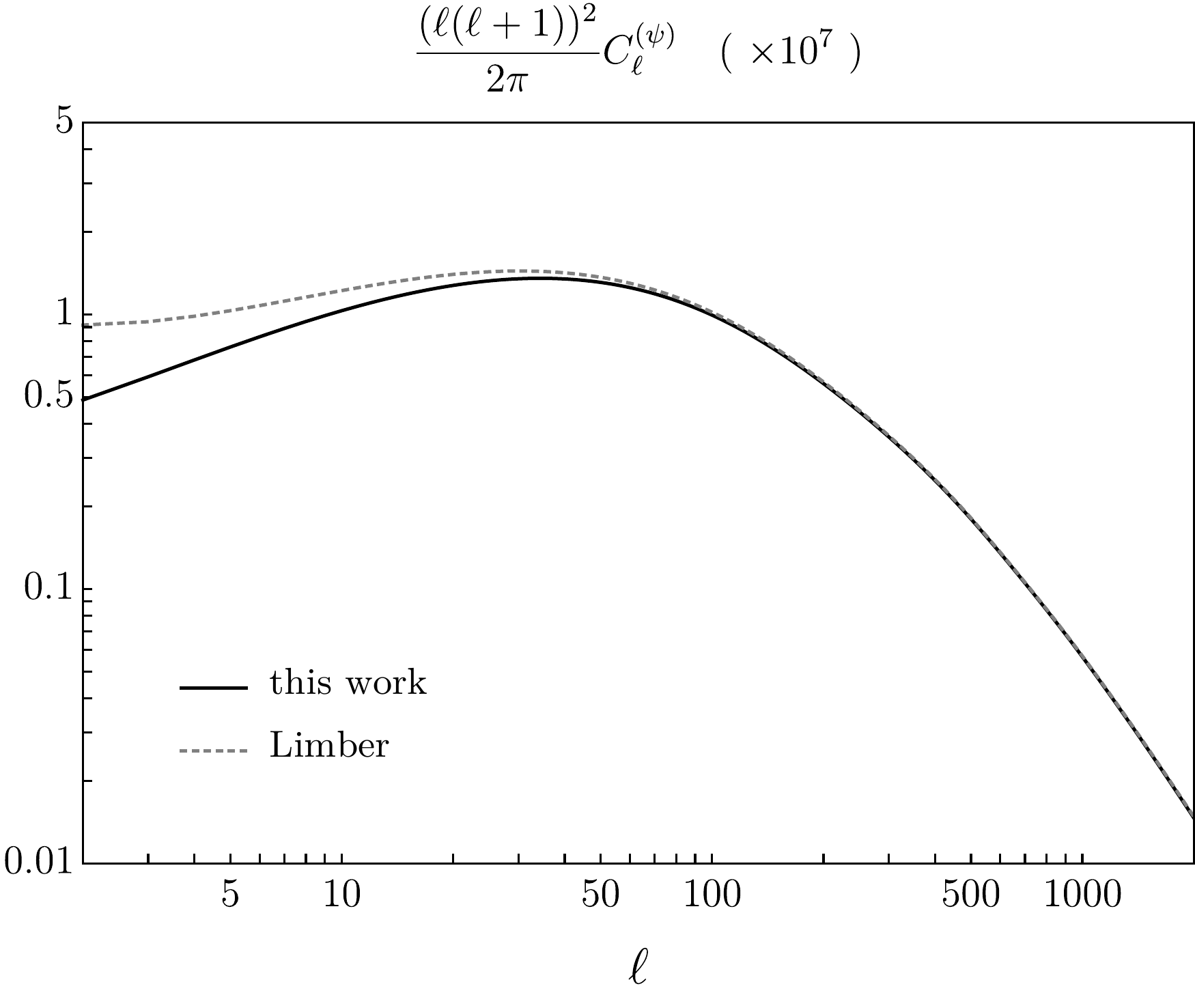}\hspace{0.3cm}
\includegraphics[scale=0.43]{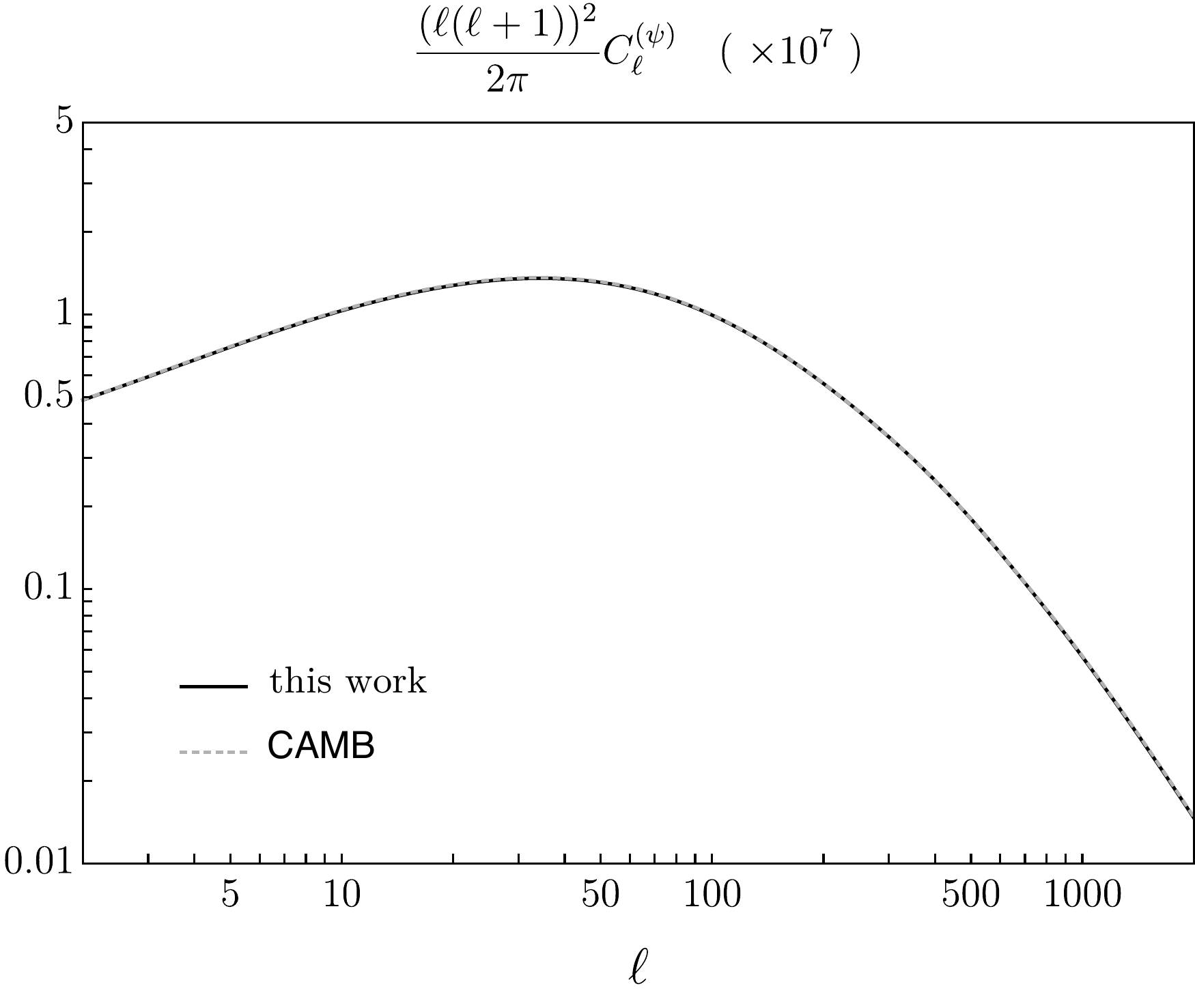}
\caption{Tree-level lensing power spectrum evaluated using~(\ref{eq:Cpsi}). {\it Left:} comparison between the result of this work (solid black line) and the Limber approximation (dashed gray line). As expected, the two results match at high multipoles ($\ell\gtrsim100$). {\it Right:} comparison between the result of this work (solid black line) and the output of {\sf CAMB} (dashed gray line). To produce these plots we used the following parameters: $N_\nu = 100$ frequencies in the FFTlog with a bias of $b=1.9$. For the line-of-sight integrals, the window functions have been expanded in Legendre polynomials up to order $31$ (which correspond to only 15 non-zero coefficients see footnote~\ref{foot:odd}). \label{fig:result}}
\end{figure}

We now turn to the lensing power spectrum. For simplicity, we assume vanishing spatial curvature and we work in the Born approximation. Under these assumptions, the lensing potential is~\cite{Lewis:2006fu}
\beq
\psi(\n)\equiv -2\int_0^\ccmb\d\chi\ W_\Phi(\chi)\Phi\big(\chi\n,z\big)\ ,
\eeq
where $\ccmb$ is the comoving distance to the last scattering surface, $\Phi$ the gravitational potential and $W_\Phi(\chi)$ is the lensing window function:
\beq
W_\Phi(\chi)\equiv \frac{\ccmb-\chi}{\ccmb\chi}\ .\label{eq:Wlens}
\eeq
Using Poisson's equation---$\Phi(\k,z)=-\frac{3}{2}\Omega_mH_0^2\,k^{-2}(1+z)\delta(\k,z)$---and working in the linear approximation in the matter overdensity $\delta$, the lensing power spectrum can be written as follows
\beq
\hspace{-6pt} C_\ell^{(\psi)}=\frac{36}{\pi} (H_0^2\Omega_m\ccmb)^2\hskip -4pt\int_0^{\ccmb}\frac{\d\chi}{\chi^2} \int_0^{\chi}\hskip -2pt\frac{\d\chi'}{(\chi')^2}\ w_\Phi\Big(\tfrac{\chi}{\ccmb}\Big) w_\Phi\Big(\tfrac{\chi'}{\ccmb}\Big)\hskip -4pt \int_0^\infty \frac{\d k}{k}\,j_{\ell}(k\chi)j_{\ell}(k\chi')[k^{-1}P_{in}(k)]\ ,\label{eq:Cellpsitree}
\eeq
where $P_{in}(k)$ is the initial matter power spectrum. The lensing window function was rescaled such that~$w_\Phi(\chi/\chi_\star)\to0$ in the limits~$\chi\to0$ and $\chi\to\ccmb$ (see fig.~\ref{fig:window}):
\beq
w_\Phi(\chi/\chi_\star)\ \equiv\ \frac{\chi^2}{\ccmb}(1+z)W_\Phi(\chi)D(\chi)\ .\label{eq:window}
\eeq
One can then proceed as in the previous section and calculate the power spectrum. However, when the window function has a broad support, there is a cute trick which significantly speeds up the computational time.
Let us be more precise: Our choice of the rescaling of the window function~(\ref{eq:window}) was motivated by the fact that it can be well approximated by a low-order polynomial (see fig.~\ref{fig:window}). Concretely, we choose to expand the window function in Legendre polynomials:
\beq
w_\Phi(u)\ =\ \sum_{p}w_p\,{\cal P}_{p}(u)\ =\ \sum_{p}\omega_p\,u^p\ ,\label{eq:Legendre}
\eeq
where ${\cal P}_{p}(u)$ is the $p$-th order Legendre polynomial and in the last line we expanded each polynomial in powers of $u$.\footnote{Interestingly, given that the window function $w_\Phi(u)$ is approximately linear in $u$ around zero, in practice we extend the window function to negative values of $u$ by imposing it to be odd, so that the sum in (\ref{eq:Legendre}) runs over odd integers only.\label{foot:odd}} 
In fig.~\ref{fig:window}, we plotted the window function along with its Legendre expansion, evaluated up to order~$p_{\rm max}=31$. In this case, the Legendre expansion matches the window function to better than $0.2\%$ accuracy. The general integral~(\ref{eq:Cell2}) then becomes
\beq
C_\ell^{(\psi)}\ =\ \frac{9}{\pi^2}\Omega_m^2H_0^4\sum_{n} c_n\chi_\star^{-\nu_n}\sum_{p_1,p_2}\omega_{p_1}\omega_{p_2}\int_0^1\d u\ u^{p_1+p_2-3-\nu_n}\int_0^1\d t\ t^{p_2-2}\,{\sf I}_\ell(\nu_n,t)\ ,\label{eq:Cpsi}
\eeq
where $u\equiv \chi/\chi_\star$. 
The $u$-integral converges only when $p_1+p_2-2-{\rm Re}(\nu_n)>0$. When this is satisfied, both the integral in $u$ and $t$ can be done analytically. In particular, the integral in $t$ is given by (see Appendix~\ref{app:IntIell})
\beq
\int_0^1\d t\ t^{p_2-2}\,{\sf I}_\ell(\nu_n,t)\ =\ \frac{\pi^{3/2} \Gamma(2 - \tfrac{\nu_n}2) \Gamma(\ell + \tfrac{\nu_n}2)}{\Gamma(\tfrac{5- \nu_n}2) \Gamma(3 + \ell - \tfrac{\nu_n}2)} \,_3F_2\left(\begin{array}{c}
1\,,\,2 + \tfrac{\ell-p_2}2 \,,\,3 - \nu_n \vspace{3pt}\\
3 + \ell - \tfrac{\nu_n}2 \, ,\tfrac{5-\nu_n}2
\end{array};\, 1\, \label{eq:IInt}
\right)\ ,
\eeq
which is very fast to evaluate, particularly in the limit $\ell\gg1$. One can immediately see the benefit of this approach: we replaced the integrals in $\chi$ and $t$, which require about $50\times 50$ sampling points to a double sum (from the Legendre expansion) which in practice requires only about $15\times 15$ terms to compute. The results are shown in fig.~\ref{fig:result}. The left plot shows a comparison between our result and the Limber approximation. As expected, the two curves match at high multipoles $\ell$. In fact, a more detailed analysis shows that, assuming full sky,  ${\rm SNR}_P\simeq 5$ when summing over the first 500 multipoles. Therefore, the two calculations yield results which are observationally distinct.  The right plot shows a comparison between our computation and the output from {\sf CAMB}~\cite{Lewis:1999bs}, which, as one can see, agree very well~(within about $1\%$ accuracy).

\subsection{CMB Anisotropies}
\label{sec:CMBPS}
\begin{figure}[h!]
\centering
\includegraphics[scale=0.55]{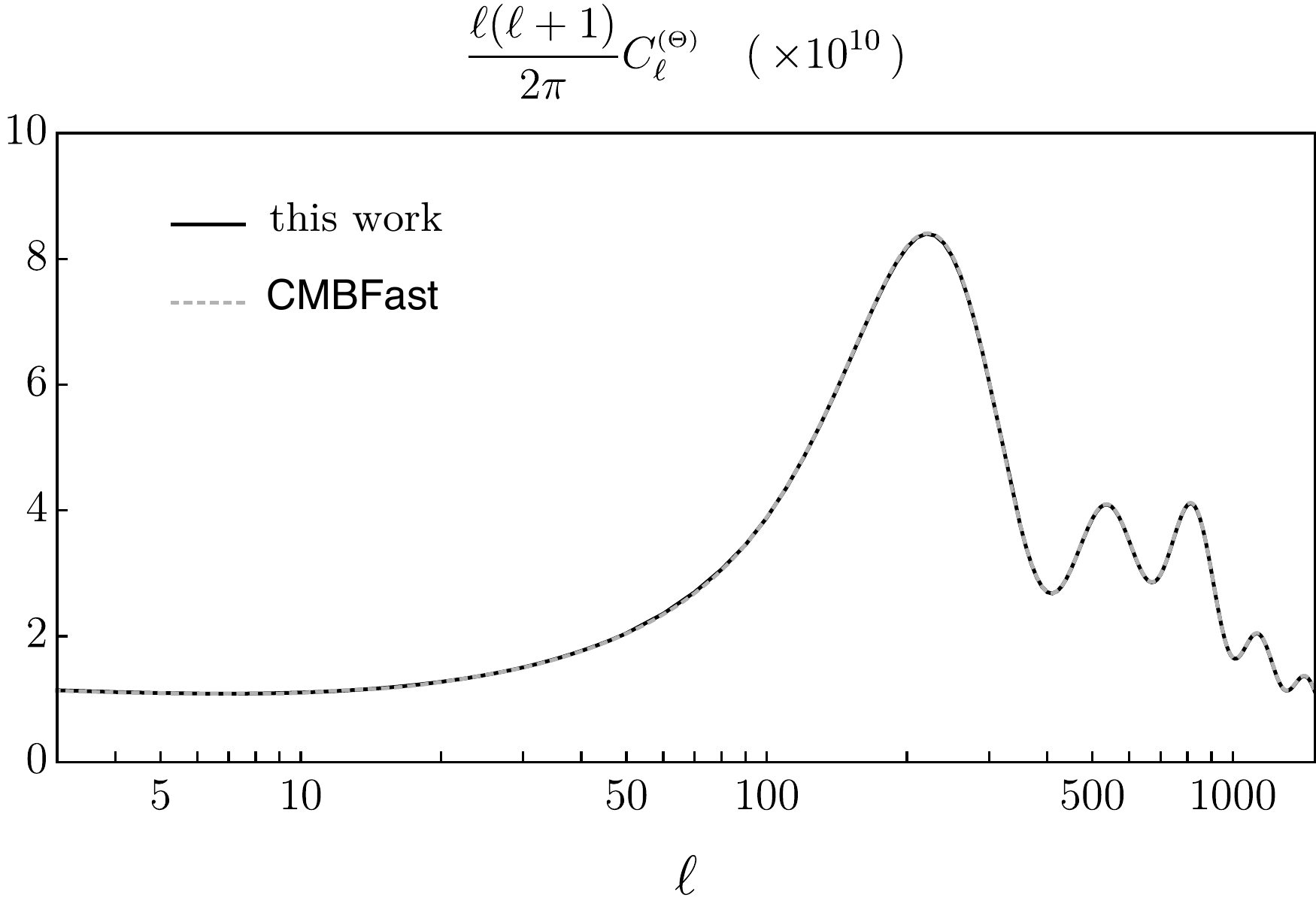}
\caption{CMB temperature angular power spectrum using our method (black solid line) and the ouput of {\sffamily CMBFast} (gray dashed line). To produce this plot, we used the following parameters:  $N_\nu = 100$ frequencies in the FFTlog with a bias of $b = 1.1$. For the line-of-sight integrals, we used $N_\chi = 60$ and $N_t=120$ sampling points.\label{fig:CMBPS}}
\end{figure}

We finally turn to the CMB temperature anisotropies $\Theta\equiv \delta T/{\bar T}$ whose power spectrum is
\beq
C_\ell^{\mathsmaller{(\Theta)}} = 4\pi\int_0^{\ccmb}\d\chi\int_0^{\ccmb}\d\chi'\int_0^\infty\frac{\d k}{k}j_\ell(k\chi)j_\ell(k\chi') [{\cal S}(k,\chi){\cal S}(k,\chi')\Delta^2_{\phi}(k)]\ ,
\eeq
where $\Delta^2_\phi\equiv\frac{k^3}{2\pi^2}P_{\phi}$ is the dimensionless power spectrum of the primordial fluctuations $\phi$ and~${\cal S}(k,\chi)$ is the CMB transfer function, which we compute using {\sffamily CMBFast}~\cite{Seljak:1996is}. There are two important differences compared to the two previous sections~\S\ref{sec:PSg} and \S\ref{sec:PSlensing}. The first one is that the transfer functions are no longer separable in space and time and we need to do a Fourier transform for each pair of points along the line of sight $(\chi,\chi')$. The second one is that~${\cal S}(k,\chi)$ has support everywhere along the line of sight and has a very sharp feature close to the last scattering surface~$\ccmb$. In practice, this means that one has to sample the transfer function around the feature with higher density of points in order to avoid numerical errors. The output of {\sffamily CMBFast} is compared with our method in fig~\ref{fig:CMBPS}. The two agree within one percent accuracy (and the agreement gets better with increasing $\ell$.) 
\vskip 4pt
Let us finally mention that {\sffamily CMBFast} takes about the same number of operations as our method. This is because in {\sffamily CMBFast} the integrals along the line of sight are performed first by computing $\Delta_\ell(k)\equiv\int\d\chi\ j_\ell(k\chi){\cal S}(k,\chi)$. Despite the fact that the transfer function $\Delta_\ell(k)$ is highly oscillatory, these oscillations do not cancel in the final integral $\int\frac{\d k}{k}[\Delta_\ell(k)]^2$. This allows the integral in~$k$ to be computed using a relatively small number of sampling points. Given that the number of operations scales differently in the two different methods, the fact that they are the same for the power spectrum is somehow a coincidence. As we will see, this is no longer the case for the CMB bispectrum, for which our method becomes more competitive.

\section{Angular Bispectrum}
\label{sec:3pt}
The angular bispectrum is famously hard to compute. In this section, using concrete examples, we show how our method provides a computationally efficient way to evaluate the three-point function. 
In Section~\ref{sec:method}, we found that the angular bispectrum of an observable~${\cal O}$~is
\beq
\vev{{\cal O}_{\ell_1m_1}{\cal O}_{\ell_2m_2}{\cal O}_{\ell_3m_3}}\ =\ {\cal G}^{\ell_1\ell_2\ell_3}_{m_1m_2m_3}\, B_{\ell_1\ell_2\ell_3}^{\mathsmaller{(\cal O)}}\ .
\eeq
The geometrical factor ${\cal G}^{\ell_1\ell_2\ell_3}_{m_1m_2m_3}$ on the r.h.s.~of this equation is the Gaunt integral:
\beq
{\cal G}^{\ell_1\ell_2\ell_3}_{m_1m_2m_3}\ \equiv\ \sqrt{\frac{(2\ell_1+1)(2\ell_2+1)(2\ell_3+1)}{4\pi}}
\ \bigg(
\begin{array}{ccc}
\;\ell_1\;&\;\ell_2\;&\;\ell_3\; \\
0&0&0
\end{array}\bigg)
\bigg(
\begin{array}{ccc}
\ell_1&\ell_2&\ell_3\\
m_1&m_2&m_3
\end{array}\bigg)\ ,\label{eq:Gaunt}
\eeq
where ${\tiny \bigg(
\begin{array}{ccc}
\ell_1&\ell_2&\ell_3\\
m_1&m_2&m_3
\end{array}\bigg)}$ is the Wigner $3j$-symbol. Furthermore, assuming that the momentum-space bispectrum is separable (see~(\ref{eq:sepB})), the angular bispectrum $B_{\ell_1\ell_2\ell_3}^{\mathsmaller{(\cal O)}}$ becomes
\beq
B_{\ell_1\ell_2\ell_3}^{\mathsmaller{(\cal O)}}\ =\ \frac{1}{(2\pi^2)^3}\int_0^\infty\d r\ r^2\ \left[I_{\ell_1}^{(1)}(r)I_{\ell_2}^{(2)}(r)I_{\ell_3}^{(3)}(r)+{\rm perms}\right]\ ,\label{eq:angB}
\eeq
where $I^{(i)}_{\ell}(r)$ was defined in (\ref{eq:CP}). We can see the benefit of this approach: the evaluation of the angular bispectrum is reduced to a one-dimensional integrals with smooth integrands. More precisely, to evaluate the bispectrum for all triangles up to $\ell_{\rm max}$ one only need to evaluate ${\cal O}(\ell_{\rm max}\times N_r)$ integrals, where $N_r$ is the number of sampling points to evaluate the integral~(\ref{eq:angB}). This is in contrast to the direct numerical computation where one has to solve a seven-dimensional integral containing six spherical Bessel functions (see e.g.~\cite{DiDio:2016gpd}).\footnote{The original seven-dimensional integral has three integrals along the line of sight, three integrals in the wavenumbers $k_i$ and one integral in $r$ coming from the delta function. The integral over $r$ can be solved analytically at the cost of coupling all momentum integrals.} For this reason, the direct numerical integration, involving the integral along the line of sight, is very computationally expensive. In the case of the CMB temperature bispectrum this computation was performed on supercomputers~(see e.g.~\cite{Fergusson:2006pr}).\footnote{Let us note that the CMB bispectrum is somewhat simpler as the integrals along the line of sight are already performed by Boltzmann solver codes such as {\sffamily CAMB} and integrals in $k$ are separable.} Regarding the galaxy and lensing bispectrum, we are not aware of any publicly available code which computes these statistics (including the integration along the line of sight).  In this section, we'll show how our approach allows to compute the galaxy (\S\ref{sec:galaxybisp}), lensing (\S\ref{sec:lensingbisp})  and CMB (\S\ref{sec:CMBbisp}) bispectrum efficiently.
%\vspace{5cm}
%We see that computing the bispectrum boils down to evaluating the same integrals $I_{\ell}(r)$ as in the power spectrum and is therefore of the same computational complexity.  In this section, we use~(\ref{eq:angB}) to compute the galaxy, lensing and CMB  bispectrum. 
\vskip 4pt
In cases where the use of the Limber approximation is appropriate, we will compare the two methods and assess whether the difference between them is statistically significant using the SNR:
\beq
({\rm SNR}_{B})^2\ \equiv\ \sum_{{\ell_1,\ell_2,\ell_3}}\frac{(B_{{\ell_1\ell_2\ell_3},\rm exact}^{\mathsmaller{({\cal O} )}}-B_{\ell_1\ell_2\ell_3,\rm limber}^{\mathsmaller{({\cal O})}})^2}{(\Delta B_{\ell_1\ell_2\ell_3}^{\mathsmaller{({\cal O})}})^2}\ ,
\eeq
where the bispectrum full-sky cosmic variance is given by
\beq
(\Delta B_{\ell_1\ell_2\ell_3}^{\mathsmaller{({\cal O})}})^2 = s_{\ell_1\ell_2\ell_3}\frac{4\pi}{(2\ell_1+1)(2\ell_2+1)(2\ell_3+1)}
\bigg(\begin{array}{ccc}
\ell_1&\ell_2&\ell_3\\
0&0&0
\end{array}\bigg)^{-2}\cdot
C^{\mathsmaller{({\cal O})}}_{\ell_1}\,C^{\mathsmaller{({\cal O})}}_{\ell_2}\,C^{\mathsmaller{({\cal O})}}_{\ell_3}\ ,
\eeq
and $s_{\ell_1\ell_2\ell_3} $ is a factor which is one when all multipoles are different, two when (only) two of them are equal and six when they are all equal.
\subsection{Galaxy Tomography}
\label{sec:galaxybisp}
\begin{figure}[h!]
\centering
\includegraphics[scale=0.43]{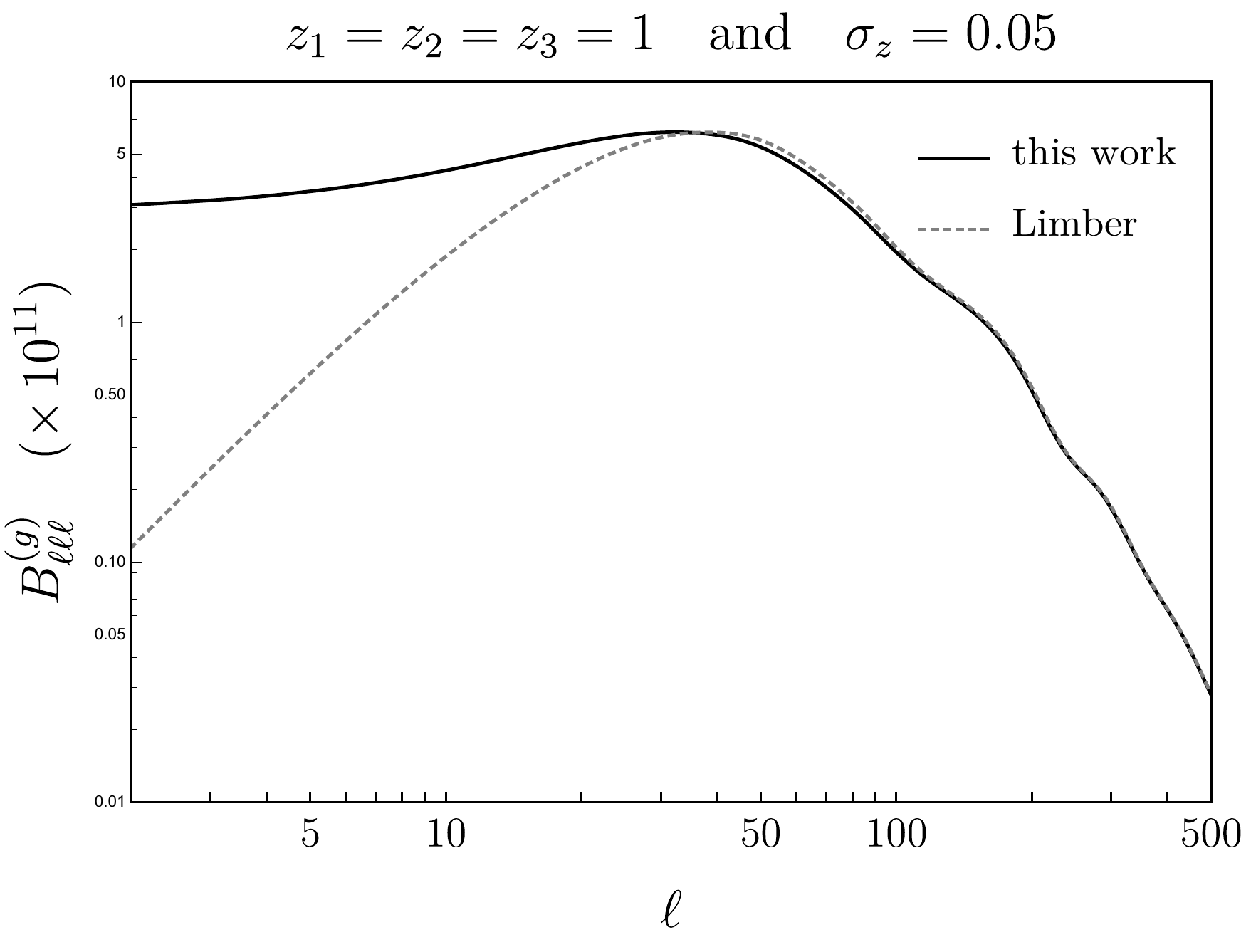}\hspace{0.3cm}
\includegraphics[scale=0.43]{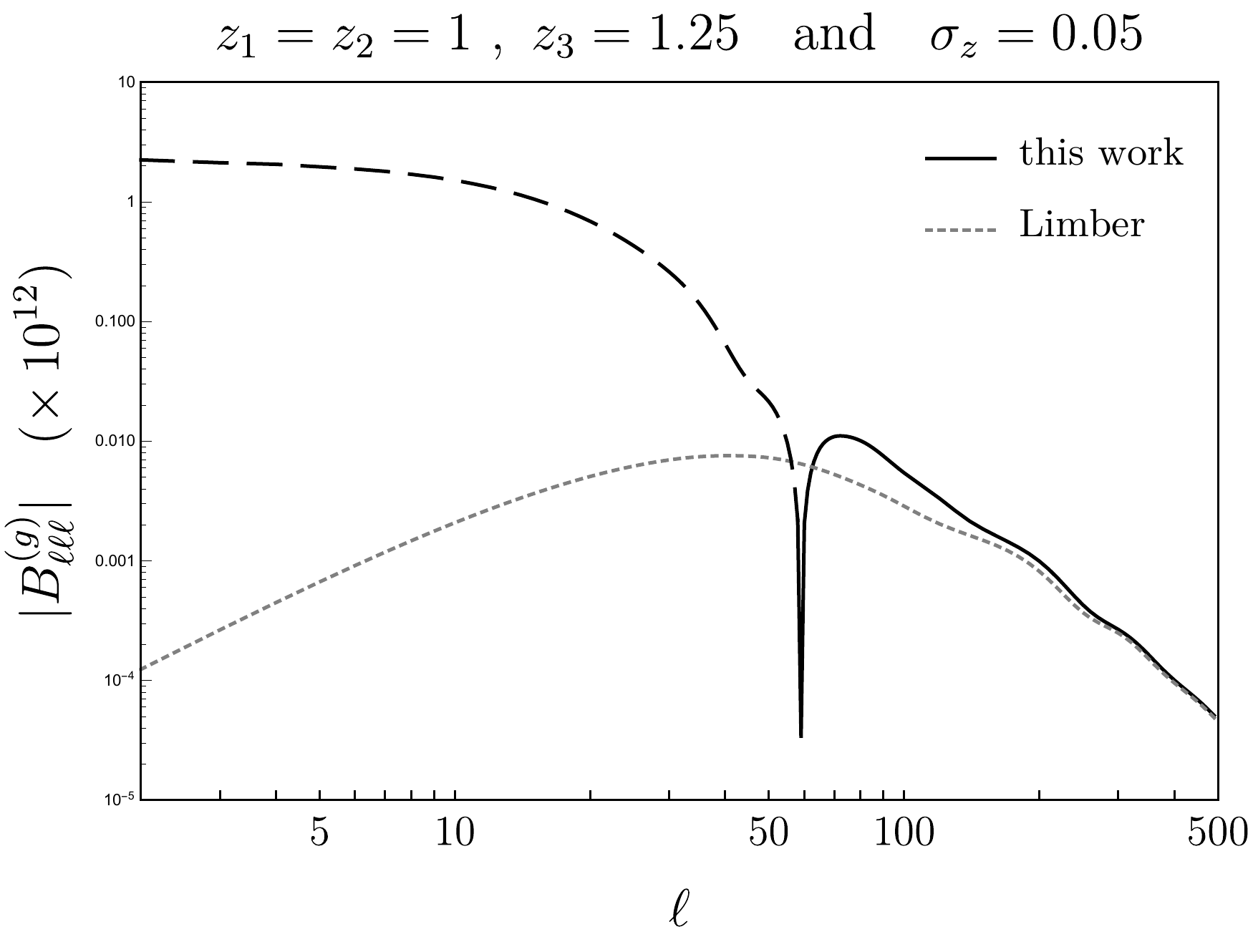}
\caption{Galaxy angular bispectrum in the equilateral configuration with $b_1=1$ and $b_2=b_{s^2}=0$ and with a redshift bin of width $\sigma_z=0.05$. {\it Left:} Correlation of three identical redshift bins ($z=1$). {\it Right:} Correlation of two identical redshift bin ($z_1=z_2=1$) with another one ($z_3=1.25$). The black dashed curve corresponds to negative values. To produce these plots, we used the following parameters:  $N_\nu = 100$ frequencies in the FFTlog with a bias of $b = 1.9$. For the line-of-sight integral and radial integral, we used $N_\chi = N_r=50$ sampling points. \label{fig:bgalaxy} }
\end{figure}

For the galaxy bispectrum, one needs to go beyond the linear bias of~(\ref{eq:linbias}) and work at second order in perturbation theory~\cite{Desjacques:2016bnm}:
\beq
\delta_g(\x,z) = b_1\,\delta(\x,z) + \tfrac{1}{2}b_2\,\delta^2(\x,z) + b_{s^2}\,s^2(\x,z)\ ,
\eeq
where $s^2\equiv (\partial_i\partial_j\Phi)^2-\tfrac{1}{3}\delta^2$ is the tidal tensor and we have again assumed redshift-independent bias parameters for simplicity. Using this expansion, one can show that, on linear scales, the galaxy bispectrum has the general form
\beq
B_g(k_i,z_i) = D_1D_2D_3^2\left[{\sf a}_0+ {\sf a}_1\,\left(\tfrac{k_2}{k_1}+\tfrac{k_1}{k_2}\right)\mu_{12} + {\sf a}_2\,\mu_{12}^2 \right]P_{in}(k_1)P_{in}(k_2) + {\rm 2\ perms}\ ,\label{eq:Bg}
\eeq
where $D_i\equiv D(z_i)$,  $\mu_{12}\equiv \hat\k_1\cdot\hat\k_2$ and the coefficients ${\sf a}_i$ depend on the bias parameters~\cite{Desjacques:2016bnm}:
\begin{align}
{\sf a_0}&\, =\, 2b_1^2\left[\frac{5}{7}b_1+\frac{1}{2}b_2-\frac{1}{3}b_{s^2}\right]\ ,\\
{\sf a_1}&\, =\, b_1^3\ ,\\
{\sf a_2}&\, =\, 2b_1^2\left[\frac{2}{7}b_1+b_{s^2}\right]\ .
\end{align}
Therefore, the galaxy bispectrum~(\ref{eq:Bg}) is a linear combination of the following terms:\footnote{We simply use momentum conservation ($\k_1+\k_2+\k_3 = \0$) to express $\mu_{12}$ in terms of the three wavenumbers $k_1$, $k_2$ and $k_3$. }
\beq
B_{n_1n_2n_3}(k_i,z_i)\ =\  \underbrace{[D_1\,k_1^{2n_1}P_{in}(k_1)]}_{\equiv\,f_{1}(k_1,z_1)}\times\underbrace{[D_2\,k_2^{2n_2}P_{in}(k_2)]}_{\equiv\,f_{2}(k_2,z_2)}\times\underbrace{[D_3^2\,k_3^{2n_3}]}_{\equiv\,f_{3}(k_3,z_3)}\ ,\label{eq:Bblock}
\eeq
where $n_{1,2}\in\{-1,0,1\}$ and $n_3\in\{0,1,2\}$. We therefore focus on the angular projection of the bispectrum~(\ref{eq:Bblock}). 
For some powers $n_i$, the momentum integrals in~(\ref{eq:CP}) are divergent.\footnote{For the integral in $k_3$, this is obvious but the convergence of the integrals in $k_1$ and $k_2$ depends on the behavior of the matter power spectrum in the UV. In our universe, the matter power spectrum does not decay sufficiently fast to allow for a direct numerical integration in the momenta~\cite{DiDio:2016gpd}.} In order to calculate the bispectrum using a direct numerical integration, one would need to do the integral in $r$ first which imposes the momenta to satisfy the triangle inequality~\cite{DiDio:2016gpd}. However, as we saw in Section~\ref{sec:method}, these divergences are spurious as they come from derivatives acting in position space (see~(\ref{eq:Helm})). All integrations which are naively divergent in $k$ can be brought to a form where they are manifestly convergent by using the identity~(\ref{eq:Ishift}).   
\vskip 4pt
First, let us focus on the integral in $k_{1}$ and $k_2$. For $i=1,2$, we have:
\begin{align}
{I}^{(i)}_{\ell}(r)&\ \equiv\  4\pi\int_0^\infty\d\chi\ \W_g^{(1)}(\chi)\int_0^\infty\frac{\d k}{k}\,j_{\ell}(k\chi)j_{\ell}(k r)\,[k^{2n_i+3}P_{in}(k)]\ \nonumber\\[4pt]
&\ =\  4\pi\int_0^\infty\d\chi\ {\cal D}_\ell^{n_i+2}[\W_g^{(1)}(\chi)]\int_0^\infty\frac{\d k}{k}\,j_{\ell}(k\chi)j_{\ell}(k r)\,[k^{-1}P_{in}(k)]\ ,\label{eq:Ig}
\end{align}
where $\W_g^{(n)}(\chi)\equiv W_{g}(\chi,\bar\chi,\sigma)[D(\chi)]^n$. 
In the second line, we repeatedly made use of~(\ref{eq:Ishift}) to reduce the powers of $k^2$ in the integrand until we reach $k^{-1}P_{in}(k)$, which is Fourier transform using~(\ref{eq:FFTlog}). In this form, ${I}^{(i)}_{\ell}(r)$ can be more easily computed (see~\S\ref{sec:PSg}).  We then turn to the integral in $k_3$ which always leads to (derivatives of) a delta function so that (\ref{eq:CP}) simplifies:
\begin{align}
{I}^{(3)}_{\ell}(r)&\ =\  4\pi\int_0^\infty\d\chi\ \W_g^{(2)}(\chi)\int_0^\infty\d k\,j_{\ell}(k\chi)j_{\ell}(k r)\,k^{2(n_3+1)}\nonumber\\[4pt]
&\ =\ \frac{2\pi^2}{r^2}{\cal D}_\ell^{n_3}[\W_g^{(2)}(\chi)]\big|_{\chi\equiv r}\ .\label{eq:Ig3}
\end{align}
Remarkably, as opposed to (\ref{eq:Ig}), {\it no integration} is required to evaluate ${I}^{(3)}_{\ell}(r)$. Once the functions~${I}^{(i)}_{\ell}(r)$  have been computed for $i\in\{1,2,3\}$,  it is easy to calculate the angular bispectrum~$B_{\ell_1\ell_2\ell_3}^{(g)}$ for all triangles using (\ref{eq:angB}).
\vskip 4pt
Notice that we use the same building blocks to compute the bispectrum as those used for the power spectrum. Furthermore,  in our method the integrals in $k$ are not only separable but can be done before integrating over $r$. As a result, the computational complexity of the bispectrum is comparable to that of the power spectrum. 
\vskip 4pt
In fig.~\ref{fig:bgalaxy}, we plotted the matter bispectrum in the equilateral configuration ($\ell_1=\ell_2=\ell_3$) for which $b_1=1$ and $b_2=b_{s^2}=0$. In this case, ${\sf a}_0 = \frac{5}{7}$, ${\sf a}_1 = \frac{1}{2}$  and ${\sf a}_2 = \frac{2}{7}$. Choosing different bias parameters will just change the values of these coefficients (but the general features of the angular bispectrum will not drastically change). We see that, just like the power spectrum, the Limber approximation gives the correct result at high~$\ell$ but fails at low multipoles. Furthermore, the difference between the exact result and the Limber approximation is more pronounced for cross-correlations. In terms of signal-to-noise, we found that, for a redshift bin of $\sigma_z=0.05$ and $z=1$, the SNR is ${\rm SNR}_{B}\simeq 20$ when summing over all triangles up to $\ell_{\rm max}=100$.
\subsection{Gravitational Lensing}
\label{sec:lensingbisp}
\begin{figure}[h!]
\centering
\includegraphics[scale=0.55]{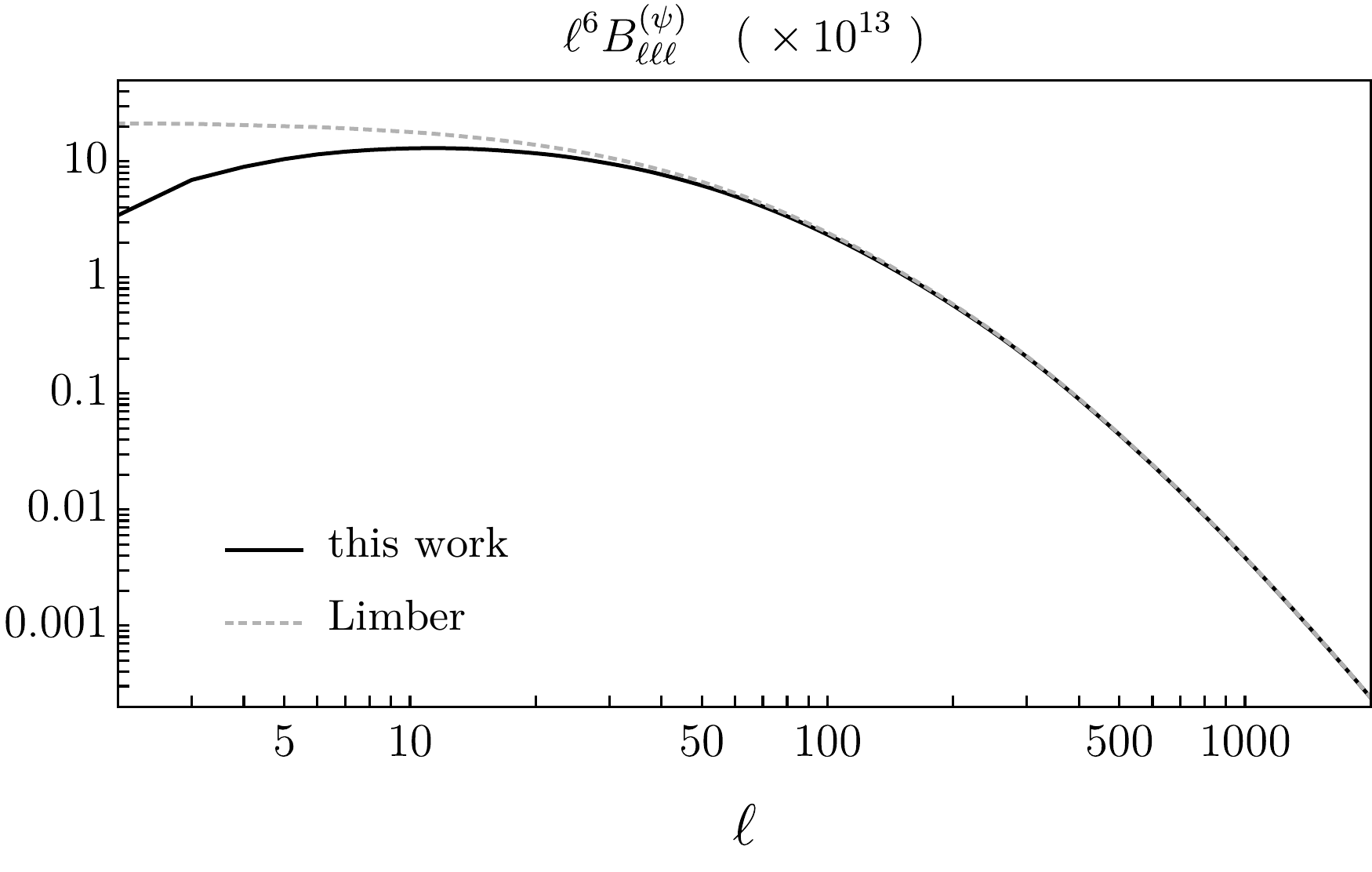}
\caption{Lensing bispectrum in the equilateral configurations using our method (black solid line) and the Limber approximation (gray dashed line), which match for $\ell\gtrsim 100$. To produce this plot, we used the following parameters:  $N_\nu = 100$ frequencies in the FFTlog with a bias of $b = 1.9$. For the line-of-sight integral and radial integral, we used $N_\chi = N_r=50$ sampling points. \label{fig:blensing}}
\end{figure}

The computation of the lensing bispectrum is similar to that of the galaxy bispectrum. The main two differences are that the observable is the gravitational potential $\Phi$ and the window function is fixed (see ~(\ref{eq:Wlens})). Given that the gravitational potential is related to the matter overdensity via Poisson's equation, we may use (\ref{eq:Bblock}) to infer that the tree-level bispectrum of the gravitational potentail can be written as a linear combination of the following terms:
\beq
B_{n_1n_2n_3}^{\mathsmaller{(\Phi)}}(k_i,z_i) = \underbrace{[(1+z_1)D_1\,k_1^{2n_1}P_{in}(k_1)]}_{\equiv\ f_1(k_1,z_1)}\times\underbrace{[(1+z_2)D_2\,k_2^{2n_2}P_{in}(k_2)]}_{\equiv\ f_2(k_2,z_2)}\times\underbrace{[(1+z_3)D_3^2\,k_3^{2n_3}]}_{\equiv\ f_3(k_3,z_3)}\ ,
\eeq
where $n_{1,2} \in\{ -2,-1,0\}$ and $n_3\in\{-1,0,1\}$. The computation of the lensing bispectrum is therefore very similar to that of the galaxy bispectrum of \S\ref{sec:galaxybisp}. There are however some technical differences which are worth pointing out:
\begin{itemize}
\item Due to the form of the lensing window function, the boundary terms in~(\ref{eq:Ishift}) do not vanish and we need to carefully keep track of them:
\begin{align}
I_\ell^{(i)}(r) & \equiv  4\pi\int_0^\infty\d\chi\ \W_{\Phi}^{(1)}(\chi)\int_0^\infty\frac{\d k}{k}\,j_{\ell}(k\chi)j_{\ell}(k r)\,[k^{2n_i+3}P_{in}(k)]\ \label{eq:Ielllensing}\\[4pt]
& =  \sum_nc_n\, r^{-(\nu_n+2(n+1))}\left[{\rm BT}(\nu_n)+\int_0^\infty\d\chi\ {\cal D}_\ell[\W^{(1)}_{\Phi}(\chi)]{\sf I_{\ell}}(\nu_n+2(n+1),\tfrac{\chi}{r})\right]\ ,\nonumber
\end{align}
where we have defined $\W^{(n)}_\Phi(\chi)\equiv W_{\Phi}(\chi)(1+z)[D(\chi)]^n$.  In the second line, we have expanded~$k^{-1}P_{in}(k)$ in Fourier modes using~(\ref{eq:FFTlog}) and~${\rm BT}(\nu_n)$ are the boundary terms:
\begin{align}
{\rm BT}(\nu)&\ \equiv\ \frac{\partial\W^{(1)}_{\Phi}(\chi)}{\partial\chi}\biggr|_{\chi\equiv\ccmb}\,{\sf I}_\ell\big(\nu+2(n+1),\tfrac{\ccmb}{r}\big)\nonumber\\
&\qquad +\ \lim_{\chi\to0}\left[\frac{2}{\chi}\W^{(1)}_\Phi(\chi)+\frac{\partial\W^{(1)}_{\Phi}(\chi)}{\partial\chi}+\W^{(1)}_\Phi(\chi)\frac{\partial}{\partial\chi}\right]{\sf I}_{\ell}\big(\nu+2(n+1),\tfrac{\chi}{r}\big)\ .
\end{align}
Notice that,  since ${\sf I}_{\ell}(\nu,t)\propto t^\ell$ as $t\to0$, the second line is non-vanishing only for $\ell=2$. 
\item When $n_3=-1$, the integral over $k_3$ doesn't yield a delta function. Instead, we have
\begin{align}
{I}^{(3)}_{\ell}(r)&\ \equiv\  4\pi\int_0^\ccmb\d\chi\ \W^{(2)}_\Phi(\chi)\int_0^\infty\d k\ j_{\ell}(k\chi)j_{\ell}(k r) k^{2(n_3+1)}\,\nonumber\\[15pt]
&\ =\ \left\{\begin{array}{ll}
\displaystyle\frac{1}{r}\int_0^\ccmb\d\chi\ \W^{(2)}_\Phi(\chi)\,{\sf I}_\ell\big(0,\tfrac{\chi}{r}\big)&\qquad{\rm when}\ n_3=-1\ ,\\[20pt]
\displaystyle\frac{2\pi^2}{r^2}{\cal D}_\ell^{n_3}[\W_{\Phi}^{(2)}(\chi)]\big|_{\chi\equiv r}\qquad&\qquad{\rm when}\ n_3\geq 0\ ,
\end{array} \right. \label{eq:Ilensing}
\end{align}
where ${\sf I}_\ell(0,t)\propto t^\ell$ for $t<1$. 
\end{itemize}
The results are shown in fig.~\ref{fig:blensing}, where our calculation is once again compared to the Limber approximation for equilateral configurations. We see that, like the lensing power spectrum, the two calculations agree for $\ell\gtrsim 100$. A more detailed analysis shows that the two approaches are not so statistically distinct, with a SNR of ${\rm SNR}_B\simeq 3$ (summing over all triangles of the first 100 multipoles). Let us finally mention that the technology developed for the lensing power spectrum (see \S\ref{sec:PSlensing}), where the window function is expanded in power laws, can also be applied to the calculation of (\ref{eq:Ielllensing}). This would, in principle, significantly speed up the calculation of the lensing bispectrum.
\subsection{CMB Anisotropies}
\label{sec:CMBbisp}
\begin{figure}[h!]
\centering
\includegraphics[scale=0.55]{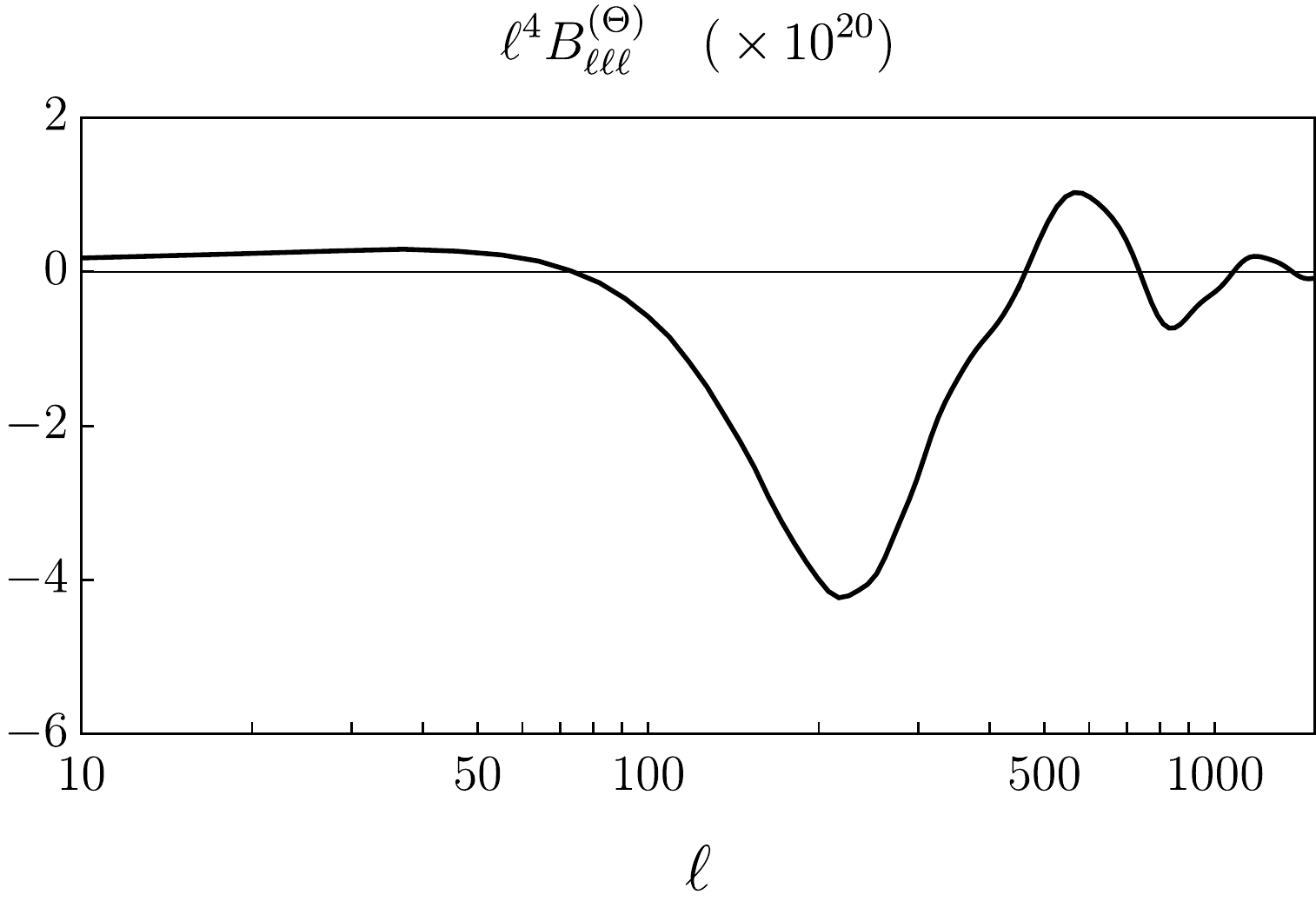}
\caption{CMB bispectrum from a flat primordial shape~(\ref{eq:flat}) in the equilateral configuration. To produce this plot, we used the following parameters:  $N_\nu = 70$ frequencies in the FFTlog with a bias of $b = 0.1$. For the line-of-sight integral and radial integral we used $N_\chi = 200$ and $N_r=80$ sampling points. (Notice that, since the last scattering surface is very thin with respect to $\ccmb$, the Limber result is never a good approximation at these scales). \label{fig:CMBBS}}
\end{figure}

We conclude our list of applications by illustrating how our method can be used to calculate the CMB temperature bispectrum. For simplicity, we present the calculation for the flat primordial shape:\footnote{We've assumed exact scale invariance for the power spectrum such that $\Delta_\phi^2$ is independent of $k$. It is however straightforward to introduce the spectral tilt.} 
\beq
B_\phi( k_1,k_2,k_3) \ =\  \frac{\Delta_\phi^4}{k_1^2k_2^2k_3^2} \ .\label{eq:flat}
\eeq
In this case, the angular bispectrum of temperature fluctuations takes a very simple form:
\beq
B_{\ell_1\ell_2\ell_3}^{(\Theta)} \ = \  \frac{\Delta_\phi^4}{(2\pi^2)^3}\int_0^\infty\d r\ r^2\left[I_{\ell_1}(r)I_{\ell_2}(r)I_{\ell_3}(r) \right]\ ,
\eeq
where 
\beq
{I}_\ell(r) \ \equiv \ 4\pi\int_0^\infty\d\chi\ \int_0^\infty\frac{\d k}{k}\,j_{\ell}(k\chi)j_\ell(k r)\,[k \, {\cal S}(k,\chi)]\ . 
\eeq
Notice that these functions are similar to the functions $b_\ell(r)$ in~\cite{Fergusson:2009nv}. Using~(\ref{eq:CP2}), the integral in $k$, which is numerically the most challenging, can be evaluated with $\mathcal O(100)$ operations. Our results are shown in fig.~\ref{fig:CMBBS}, where we plotted the bispectrum for equilateral triangles as a function of the multipole $\ell$. Just like the CMB power spectrum, we use {\sffamily CMBFast} to compute the transfer function ${\cal S}(k,\chi)$. The angular bispectrum has the characteristic oscillatory features and our result is in agreement with previous calculations~\cite{Fergusson:2006pr}.
\vskip 4pt
Finally, let us mention that it should be simple to apply our method to other separable shapes with different momentum dependence or other observables such as $E$- or $B$-modes of the CMB polarization. In some applications, it is useful to reduce powers of momenta using identity~(\ref{eq:Ishift}). For example, that would be the case for one of the momentum integrals if we were to consider local non-Gaussianity.  When dealing with CMB this requires extra care because ${\cal S}(k,\chi)$ is always evaluated numerically. The problem lies in the fact that even very small numerical error in sampling points can cause problems in evaluating derivatives in $\chi$ if one uses simple interpolating schemes. However, this is not a fundamental limitation and numerical issues can be avoided by increasing the precision, changing the format of the output of Boltzmann codes or using numerical algorithms for calculating derivatives which are insensitive to small random noise in the data points. 

\section{Discussion}
In this paper, we presented a new efficient method to numerically evaluate cosmological angular statistics. The main bottleneck to computing the angular power spectrum and bispectrum comes from the integrals of rapidly-oscillating spherical Bessel functions (see~(\ref{eq:CP})). We showed that, by projecting the momentum-space statistics on a basis of power-law functions (using e.g.~an FFTlog algorithm) these integrals can be evaluated using only about 100 operations. Remarkably, this number of operations does not change with the value of the multipole $\ell$.
\vskip 4pt
Our approach relies mainly on the condition that the power spectrum and bispectrum are separable in the momenta. This is however not as stringent as it seems. First, we saw that this assumption can be easily circumvented for the power spectrum (see~(\ref{eq:Cell2})). For the bispectrum, this assumption cannot be circumvented, but we showed that the separability condition is often met in practice. For instance, the galaxy and lensing bispectrum are separable on large scales, which is precisely the regime where the Limber or flat-sky approximations fail and  an exact calculation is necessary. On smaller scales, this condition is no longer satisfied, but in this regime approximate methods provide reasonably accurate results. Finally, the CMB bispectrum is in general not separable. However, it was shown in~\cite{Fergusson:2008ra} that the primordial bispectrum can be expanded onto a basis of separable shapes. Naturally, in principle, one could also Fourier transform a non-separable bispectrum and apply the same method as the one developed in this paper. However, such an approach is likely to be difficult since $(i)$~the 3D Fourier transform of the bispectrum may be computationally expensive and $(ii)$~the complexity of the calculation would grow as ${\cal O}(\ell_{\rm max}^3)$.
\vskip 4pt
In this work, we only considered scalar (spin-0) observables. However, spin-1 (or higher) quantities---such as CMB polarization or shear weak lensing---are also of interest in cosmology. In this case, one still needs to solve the same momentum integral as in~(\ref{eq:CP}) (see e.g.~\cite{Kilbinger:2017lvu,Bernardeau:2011tc}) and our general approach straightforwardly applies to these observables. 
\vskip 4pt
Finally, let us conclude by mentioning that the method presented in this work is particularly useful to accurately compute large-scale effects. We illustrated this point by showing that the scale-dependent bias induced by primordial non-Gaussianities and redshift-space distortions are indeed poorly captured by the Limber approximation (see \S\ref{sec:PSg}). There are however many other effects which are important on large scales, such as e.g.~relativistic corrections. It would be interesting to calculate these effects using our method.

\subsubsection*{Acknowledgements}
We thank James Fergusson, Marcel Schmittfull, Paul Shellard and Gabriele Trevisan for discussions, as well as Enea Di Dio for helpful comments on the manuscript. We are also grateful to Jean-Baptiste Fouvry for his help with {\sffamily Mathematica}. V.A. gratefully  acknowledges support from the Ralph E.~and Doris M.~Hansmann Membership. M.S. gratefully acknowledges support from the Institute for Advanced Study and the Raymond and Beverly Sackler Foundation. M.Z. is supported by the NSF grants PHY-1213563, AST-1409709, and PHY-1521097.

\newpage

\appendix

\section{Hypergeometric Functions}
\label{app:hyper}
In this appendix, we provide definitions of hypergeometric functions as well some of their properties which are used in the main text. 
\vskip 10pt
\noindent{\it Hypergeometric function.---}The hypergeometric function $\,_2F_1(a,b,c,z)$ is a solution of Euler's hypergeometric equation:
\beq
z(1-z)\,f''(z)+\big(c-(a+b+1)z\big)\,f'(z) -ab\,f(z) = 0\;,
\eeq
where $a$, $b$ and $c$ are arbitrary complex numbers. In the region $|z|<1$, the hypergeometric function has the following power series representation:
\beq
\,_2F_1(a,b,c,z) \ = \ \frac{\Gamma(c)}{\Gamma(a)\Gamma(b)}\sum_{n=0}^\infty \frac{\Gamma(a+n)\Gamma(b+n)}{\Gamma(c+n)n!} z^n \;,
\label{eq:2F1_power}
\eeq
which can be used for numerical evaluation. The series converges for $z=1$ only when the parameters satisfy~${\rm Re}(c-a-b)>0$. In this case, the hypergeometric function has a simple expression in terms of gamma functions:
\beq
\,_2F_1(a,b,c,1) \ = \ \frac{\Gamma(c)\Gamma(c-a-b)}{\Gamma(c-a)\Gamma(c- b)} \;, \quad {\rm Re}(c-a-b)>0\;.
\eeq
Even though the hypergeometric series~(\ref{eq:2F1_power}) is formally convergent everywhere in the unit disk, the convergence for arguments close to $|z|=1$ can be very slow in practice. This happens whenever the absolute value of either $a$ or $b$ (or both) is significantly larger than the absolute value of $c$. In such situations not only one has to calculate many terms in the series to achieve convergence, but also these terms need to be evaluated with a very high number of significant digits due to fine cancellations among very large numbers.
A way to solve this problem is to use functional identities that map points close to $|z|=1$ to a region around $z=0$ where the series converges rapidly. One such identity is 
\begin{align}
\,_2F_1(a,b,c,1-z) &\ =  \ \frac{\Gamma(c) \Gamma(c - a - b)}{\Gamma(c - a) \Gamma(c - b)} \,_2F_1(a, b, a + b - c + 1, z) \nonumber \\
& \qquad +\ \frac{\Gamma(c) \Gamma(a + b - c)}{\Gamma(a) \Gamma(b)} z^{c - a - b} \,_2F_1(c - a, c - b, 1 - a - b + c, z) \;.
\label{eq:zto(1-z)}
\end{align}
Outside the region $|z|<1$, the hypergeometric function can be calculated using analytical continuation. A useful formula to keep in mind is
\beq
{}_2F_1(a,b,c,1/z)\ =\  \frac{\Gamma(b - a) \Gamma(c)}{\Gamma(b) \Gamma(c - a)} (-z)^a\,_2F_1(a, a - c + 1, a - b + 1, z) \; + \; (a\leftrightarrow b)\;.
\label{eq:zto1/z}
\eeq
These two identities are the basic transformations needed to evaluate the hypergeometric function in the entire complex plane. For some specific values of the parameters $(a,b,c)$, there are additional special transformations which can speed up the evaluation of the hypergeometric function, such as, for instance, the following quadratic transformation:
\beq
\,_2F_1(a, b, 2 b, 1-z^2) \, =\, z^{-a} \,_2F_1\left(\tfrac{a}{2}, b - \tfrac{a}{2}, b + \tfrac{1}{2}, -\tfrac{(1-z^2)^2}{4 z^2}\right) \;.
\label{eq:2F1_quadratic_transf}
\eeq
Another interesting property of hypergeometric function is that any function of the form
\beq
\,_2F_1(a+n_1, b+n_2, c+n_3, z)\;,
\eeq
where $n_i\in\mathbb{Z}$, can always be expressed in terms of a linear combination of ${}_2F_1(a, b, c, z)$ and any contiguous function\footnote{A contiguous function is a function in which any of the parameters is shifted by $+1$ or $-1$, such as e.g.~${}_2F_1(a+1, b, c, z)$}:
\beq
\,_2F_1(a+n_1, b+n_2, c+n_3, z) \ = \ \alpha\times{}_2F_1(a, b, c, z) +\beta\times{}_2F_1(a+1, b, c, z) \;.
\eeq
The coefficients $\alpha$ and $\beta$ are some rational functions of $a$, $b$, $c$ and $z$. This property can be used to find recursion relations, which can additionally speed up calculations of hypergeometric functions. For example, one such relation is 
\beq
{}_2F_1(a, b + 2, c + 2, z) = \tfrac{c(c +1)}{(b + 1)(c - a +1)z} \left[(1 - \tfrac{a - b - 1}c z) {}_2F_1(a, b + 1, c + 1, z) - {}_2F_1(a, b, c, z) \right] \;.
\label{eq:2F1rec}
\eeq
\vskip 10pt
\noindent{\it Generalized hypergeometric functions.---}The basic hypergeometric series~(\ref{eq:2F1_power}) can be extended to define generalized hypergeometric functions:
\beq
{}_pF_q\left(\begin{array}{c}
a_1\,,\,a_2\,,\ldots,\,a_p\\
b_1\,,\,b_2\,,\ldots,b_q
\end{array};\, z\, 
\right) \ \equiv \ \frac{\Gamma(b_1)\cdots\Gamma(b_q)}{\Gamma(a_1)\cdots \Gamma(a_p)}\sum_{n=0}^\infty \frac{\Gamma(a_1+n)\cdots \Gamma(a_p+n)}{\Gamma(b_1+n)\cdots \Gamma(b_q+n)} \frac{z^n}{n!} \;,
\label{eq:pFq_power}
\eeq
where $p$ and $q$ are positive integers.  When $p=q+1$---which is the most commonly encountered case---the generalized hypergeometric series~(\ref{eq:pFq_power}) converges for $|z|<1$. Outside the unit disk, the generalized hypergeometric function can be calculated using analytical continuation. A special point is $z=1$ where the series converges only when~${\rm Re}(b_1+\cdots+b_q-a_1-\cdots a_{q+1})>0$. One important identity that we use for the lensing power spectrum is 
\beq
{}_3F_2\left(\begin{array}{c}
a_1\,,\,a_2\,,\,a_3\\
b_1\,,\,b_2
\end{array};\, 1\, 
\right) \ = \ {\cal N}\times {}_3F_2\left(\begin{array}{c}
b_1 - a_1\,,\,b_2 - a_1\,,\,b_1 + b_2 - a_1 - a_2 - a_3\\
b_1 + b_2 - a_1 - a_2\,,\,b_1 + b_2 - a_1 - a_3
\end{array};\, 1\, 
\right) \;,
\label{eq:3F2_transf}
\eeq
where 
\beq
{\cal N} \ \equiv \ \frac{\Gamma(b_1)\Gamma(b_2) \Gamma(b_1+b_2 - a_1-a_2-a_3)}{\Gamma(a_1) \Gamma(b_1 + b_2 - a_1 - a_2) \Gamma(b_1 + b_2 - a_1 - a_3)} \;.
\eeq
\vskip 4pt
In this appendix, we have listed all identities that are used in this paper to evaluate  angular power spectra and bispectra. The theory of hypergeometric functions is very rich and there are many other properties~\cite{book1,book2} that are potentially useful in this context.

\section{Efficient Evaluation of $\boldsymbol{{\sf I}_{\ell}(\nu,t)}$}
\label{app:Iell}
We want to find an optimal way to calculate ${\sf I}_{\ell}(\nu,t)$:
\beq
{\sf I}_{\ell}(\nu,t) = 4\pi\int_0^\infty \d v\; v^{\nu-1}j_\ell(v) j_\ell (v t) \;.\label{eq:Idef}
\eeq
A direct integration gives
\beq
{\sf I}_\ell(\nu,t)\ =\  \frac{2^{\nu-1}\pi^2\,\Gamma(\ell+\tfrac{\nu}{2})}{\Gamma(\tfrac{3-\nu}{2})\Gamma(\ell+\tfrac{3}{2})}\;
t^\ell\;_2F_1\left(\tfrac{\nu-1}{2},\ell+\tfrac{\nu}{2},\ell+\tfrac{3}{2},t^2\right) \quad {\rm for\ } \quad t<1 \;.
\label{eq:I_t<t*}
\eeq
When $t>1$, it is easy to see from (\ref{eq:Idef}) that this function satisfies 
\beq
{\sf I}_{\ell}(\nu,t) = t^{-\nu} {\sf I}_{\ell}(\nu,1/t) \ .
\eeq
Complex gamma functions can be evaluated using the Lanczos approximation~\cite{Press1996}. 
The task therefore boils down to efficiently evaluate $\;_2F_1\left(\tfrac{\nu-1}{2},\ell+\tfrac{\nu}{2},\ell+\tfrac{3}{2},t^2\right)$ for $t<1$. In principle this can be done using power series expansion~(\ref{eq:2F1_power}), but for large $\ell$ and $\nu$ this can be very slow (particularly close to $t=1$) and numerically unstable. The strategy then is to use~(\ref{eq:2F1_power}) whenever $t<t_\star$ where $t_\star$ is chosen such that for any relevant values of $\nu$ and $\ell$ the power series rapidly converges. In practice we choose $t_\star=0.7$. When $t>t_\star$, before evaluating the hypergeometric function, we apply the transformation~(\ref{eq:zto(1-z)}):
\begin{align}
\hskip -11pt {}_2F_1\left(\tfrac{\nu-1}{2},\tfrac{2\ell+\nu}{2},\tfrac{2\ell+3}{2},t^2\right)&   =  \frac{\Gamma(\tfrac{2\ell+3}{2}) \Gamma(2 - \nu)}{\Gamma(\tfrac{2\ell-\nu+4}{2}) \Gamma(\tfrac{3-\nu}{2})} \,_2F_1(\tfrac{2\ell+\nu}{2}, \tfrac{\nu-1}{2}, \nu-1, 1-t^2) \nonumber \\
& \quad  +\ \frac{\Gamma(\tfrac{2\ell+3}{2}) \Gamma(\nu-2)}{\Gamma(\tfrac{2\ell+\nu}{2})\Gamma(\tfrac{\nu-1}{2}) } (1-t^2)^{2 - \nu} \,_2F_1(\tfrac{2\ell-\nu+4}{2}, \tfrac{3-\nu}{2}, 3 - \nu , 1-t^2) \;.\label{eq:B4}
\end{align}
Notice that both hypergeometric functions on the r.h.s.~are of the form~$\,_2F_1(a,b,2b,1-t^2)$. We can therefore use the quadratic transformation~(\ref{eq:2F1_quadratic_transf}):
\begin{align}
 \,_2F_1\left(\tfrac{\nu-1}{2},\tfrac{2\ell+\nu}{2},\tfrac{2\ell+3}{2},t^2\right)& =  \frac{\Gamma(\tfrac{2\ell+3}{2}) \Gamma(\mathsmaller{\nu-2})}{\Gamma(\tfrac{2\ell+\nu}{2})\Gamma(\tfrac{\nu-1}{2}) } (1-t^2)^{2 - \nu} t^{\tfrac{\nu-2\ell-4}{2}}\,_2F_1\left (\tfrac{2\ell-\nu+4}4, \tfrac{2-2\ell-\nu}{4}, \tfrac{4-\nu}2 , -\tfrac{(1-t^2)^2}{4 t^2} \right)
\nonumber\\[10pt]
&\hskip10pt+\frac{\Gamma(\tfrac{2\ell+3}{2}) \Gamma(\mathsmaller{2 - \nu})}{\Gamma(\tfrac{2\ell-\nu+4}{2}) \Gamma(\tfrac{3-\nu}{2})} t^{-\tfrac{2\ell+\nu}2 }\,_2F_1\left( \tfrac{2\ell+\nu}4, \tfrac{\nu-2\ell-2}{4}, \tfrac{\nu}{2}, -\tfrac{(1-t^2)^2}{4 t^2} \right)
\end{align}
In this expression, the parameters in the hypergeometric functions are roughly twice as small as those in~(\ref{eq:B4}), since all multipole numbers $\ell$ and frequencies $\nu$ are divided by at least a factor of 2. Moreover, the argument $\tfrac{(1-t^2)^2}{4 t^2}$ is always significantly smaller than $1-t^2$ for $t\in[t_\star,1]$. Putting everything together and simplifying gamma functions, we find that in the range $[t_\star,1]$, the function ${\sf I}_\ell(\nu,t)$ can be written as follow:
\begin{align}
{\sf I}_\ell(\nu,t) \ &= \ \frac{\pi^{3/2} t^{-\tfrac{\nu}2 }}{\Gamma(\tfrac{3-\nu}{2})} \left[ \frac{\Gamma(\ell+\tfrac{\nu}{2})}{\Gamma(\ell+2 -\tfrac{\nu}{2})} \frac{\Gamma (1-\tfrac{\nu}2) }{\Gamma(\tfrac{\nu}2-1)}  \,_2F_1\left( \tfrac{\ell}2+\tfrac{\nu}{4}, \tfrac{\nu}{4}-\tfrac{\ell+1}2, \tfrac{\nu}{2}, \tilde t  \hskip 2pt\right) \right. \nonumber \\
& \left. \hskip 60pt+\ \left(-\tilde t/4 \right)^{1-\tfrac{\nu}2} \,_2F_1\left (\tfrac{\ell}2 - \tfrac{\nu}{4}+1, \tfrac{1}{2} -\tfrac{\ell}2 - \tfrac{\nu}{4}, 2 - \tfrac{\nu}2 , \tilde t \hskip 2pt \right) \right]\ ,
\label{eq:I_t>t*}
\end{align}
where $\tilde t \equiv-\tfrac{(1-t^2)^2}{4t^2}$. Equations~(\ref{eq:I_t<t*}) and~(\ref{eq:I_t>t*}) are the most efficient expressions we found for evaluating ${\sf I}_\ell(\nu,t)$ in the range $0\leq t\leq 1$. However, for very large values of $\ell$, numerical instabilities around $t=t_\star$ can still arise, regardless of whether~(\ref{eq:I_t<t*}) or~(\ref{eq:I_t>t*}) is used. Luckily, we do not have to deal with this problem in practice. For high $\ell$, the function ${\sf I}_{\ell}(\nu,t)$ sharply peaks around $t=1$ (see fig.~\ref{fig:Iell} in the main text). This comes from the $t^\ell$ factor which multiplies the hypergeometric function in~(\ref{eq:I_t<t*}). Consequently, with increasing multipole number $\ell$ the region where the function has support is getting narrower and the function sharply decays everywhere outside this region. This means that we can set ${\sf I}_{\ell}(\nu,t)=0$ for $t<t_{\rm min}(\ell,\nu)$, where  $t_{\rm min}(\ell,\nu)$ is defined such that $|{\sf I}_{\ell}(\nu,t_{\rm min}(\ell,\nu))|=\epsilon|{\sf I}_{\ell}(\nu,1)|$ for some small number $\epsilon$ (in practice we choose $\epsilon=10^{-5}$). The function $t_{\rm min}(\ell,\nu)$ is very smooth: it can be sampled once with a small number of points in $(\ell,\nu)$ and interpolated. 
\vskip 4pt
Let us finish by mentioning another useful property of ${\sf I}_{\ell}(\nu,t)$. Using recursion relations for hypergeometric functions, such as~(\ref{eq:2F1rec}), one can derive the following identity:
\beq
(3+\ell-\tfrac{\nu}2)\,{\sf I}_{\ell+2}(\nu,t) \ = \ \tfrac{1+t^2}{t} (\ell+\tfrac 32)\, {\sf I}_{\ell+1}(\nu,t) - (\ell+\tfrac{\nu}2)\, {\sf I}_{\ell}(\nu,t) \;.\label{eq:recursionIl}
\eeq
This relation allows for a very fast evaluation of all multipoles $\ell$, for a given $\nu$ and $t$. It can be used forward, starting from $\ell=0$ and $\ell=1$ for which simple formulas exist
\begin{align}
{\sf I}_0(\nu,t)&\ =\ 2 \pi  \cos \left(\tfrac{\pi  \nu }{2}\right) \Gamma (\nu -2)\,t^{-1}
 \left[(1+t)^{2-\nu}-(1-t)^{2-\nu}\right]\ ,\label{eq:I0nut}\\
{\sf I}_1(\nu,t)&\ =\ \frac{2 \pi  \cos \left(\frac{\pi  \nu }{2}\right) \Gamma (\nu -2)}{(4-\nu) t^2 }
 \left[(1+t)^{2-\nu}\left((1-t)^2+ \nu t\right)-(1-t)^{2-\nu} \left((1+t)^2-\nu 
   t\right)\right]\ .\label{eq:I1nut}
\end{align}
For fixed values of $\nu$ and $t<1$, as one goes to high values of the multipole $\ell$, the function $|{\sf I}_{\ell}(\nu,t)|$ rapidly decays as $t^{\ell}$. On the other hand, small round-off errors made in evaluating the initial conditions (\ref{eq:I0nut}) and (\ref{eq:I1nut}) grow as $t^{-\ell}$. Eventually, the error will become as large as the function ${\sf I}_\ell(\nu,t)$ itself and this procedure breaks down. In practice, this crossover happens when the function already satisfies $|{\sf I}_\ell(\nu,t)| \leq 10^{-5} |{\sf I}_\ell(\nu,1)|$ and therefore can be set to zero as explained previously. If one requires higher precision, one can re-initiate the recursion relation every few hundreds multipoles.
% In practice, recursion relations should be used with care, because small numerical errors in the initial conditions can easily proliferate.

\section{Efficient Evaluation of $\boldsymbol{\int_0^1 \d t \, t^p \, {\sf I}_{\ell}(\nu,t)}$}
\label{app:IntIell}
In \S\ref{sec:PSlensing}, integrals of ${\sf I}_{\ell}(\nu,t)$ against power law functions $t^p$ appear (see~(\ref{eq:IInt})). Here we give an analytical formula for this type of integrals. A direct integration leads to 
\beq
\int_0^1\d t\ t^{p}\,{\sf I}_\ell(\nu_n,t)\ =\ \frac{2^{\nu-1}\pi^2\Gamma(\ell+\frac{\nu}{2})}{(p+\ell+1)\Gamma(\frac{3-\nu}{2})\Gamma(\ell+\frac{3}{2})} 
\,_3F_2\left(\begin{array}{c}
\frac{\ell+p+1}{2}\,,\,\frac{\nu-1}{2}\,,\,\ell+\frac{\nu}{2} \vspace{3pt}\\
\frac{\ell+p+3}{2}\, ,\ell+\frac{3}{2}
\end{array};\, 1\, 
\right)\ .
\eeq
The generalized hypergeometric function on the r.h.s.~is very difficult to evaluate using power series, in particular for high $\ell$ and $\nu$. Luckily, using the transformation~(\ref{eq:3F2_transf}), this expression can be brought to the following form:
\beq
\int_0^1\d t\ t^{p}\,{\sf I}_\ell(\nu_n,t)\ =\ \frac{\pi^{3/2} \Gamma(2 - \tfrac{\nu}2) \Gamma(\ell + \tfrac{\nu}2)}{\Gamma(\tfrac{5- \nu}2) \Gamma(3 + \ell - \tfrac{\nu}2)} \,_3F_2\left(\begin{array}{c}
1\,,\,1 + \tfrac{\ell-p}2 \,,\,3 - \nu \vspace{3pt}\\
3 + \ell - \tfrac{\nu}2 \, ,\tfrac{5-\nu}2
\end{array};\, 1\, 
\right)\ ,
\eeq
which is much more suitable for numerical evaluation. The reason is that the sum of the first three parameters is always smaller than the sum of the last two and this difference increases with~$\ell$. Following the notation of~(\ref{eq:pFq_power}), we find
\beq
b_1+b_2 -a_1- a_2-a_3\ =\ \tfrac {\ell+p+1}2\ >\ 0 \ .
\eeq
The bigger this number is, the faster is the convergence of generalized hypergeometric series. At high $\ell$ only a few first terms in the power series are needed to reach very high precision. 
\section{Parameters and Performance}
In this section, we provide the parameters used to produce the plots of this paper in Table~\ref{Table:Cl} (for the angular power spectra) and Table~\ref{Table:Bl} (for the angular bispectra). The parameters have been chosen such that, if we were to increase the number of sampling points and frequencies, the final result would remain unchanged. In practice, one can probably use smaller parameters to reach a satisfactory precision. In these tables, we also provide the time observed on a laptop using our {\sf Mathematica} code (without parallelization). As we have already emphasized in the main text, the integration along the line of sight is not optimized in any of these examples. In a code where these integrations are optimized and recursion relations are included we expect the performance to be better.

\vspace{0.5cm}

	 \begin{table}[h!]

	\heavyrulewidth=.08em
	\lightrulewidth=.05em
	\cmidrulewidth=.03em
	\belowrulesep=.65ex
	\belowbottomsep=0pt
	\aboverulesep=.4ex
	\abovetopsep=0pt
	\cmidrulesep=\doublerulesep
	\cmidrulekern=.5em
	\defaultaddspace=.5em
	\renewcommand{\arraystretch}{1.6}

	\begin{center}
		\small
		\begin{tabular}{lcccccc}

			\toprule
	\rowcolor[gray]{0.9}{}Observable & $N_\nu$ & $b$ & $N_\chi$ & $N_t$ & $N_\ell$ & Time \\[2pt]
		\midrule
		Galaxy tomography [$C_\ell^{(g)}$]			&100 & 1.9  & 50 & 50 &  200 & 30\,s \\
		CMB lensing [$C_\ell^{(\psi)}$]		&100 &1.9& 15$^{(\star)}$      &  15$^{(\star)}$ &  200 & 5\,s \\
		CMB anisotropies [$C_\ell^{(\Theta)}$]			&100 & 1.1 & 60 & 120 &  200 &1\,min \\
		\bottomrule
		\end{tabular}
	\end{center}
	\vspace{-0.2cm}
	\caption{Parameters used to compute the power spectra. $N_{\nu}$ is the number of frequencies used in the FFTlog, $b$ is the bias, $N_{\chi}$ and $N_t$ are the number of sampling points in the $\chi$ and $t$ integral, respectively and $N_\ell$ is the number of multipoles computed. The last column corresponds to the time observed on a laptop using our {\sf Mathematica} code (without parallelization).\\{\footnotesize $^{(\star)}$\,The values of $N_\chi$ and $N_t$ refer to the order of the Legendre expansion of the window functions (see~(\ref{eq:Legendre})).}}
	\label{Table:Cl}
\end{table}

\vspace{0.5cm}

	 \begin{table}[h!]

	\heavyrulewidth=.08em
	\lightrulewidth=.05em
	\cmidrulewidth=.03em
	\belowrulesep=.65ex
	\belowbottomsep=0pt
	\aboverulesep=.4ex
	\abovetopsep=0pt
	\cmidrulesep=\doublerulesep
	\cmidrulekern=.5em
	\defaultaddspace=.5em
	\renewcommand{\arraystretch}{1.6}

	\begin{center}
		\small
		\begin{tabular}{lcccccc}

			\toprule
	\rowcolor[gray]{0.9}{}Observable & $N_\nu$ & $b$ & $N_\chi$ & $N_r$ & $N_{\ell_1\ell_2\ell_3}$ & Time \\[2pt]
		\midrule
		Galaxy tomography [$B_{\ell_1\ell_2\ell_3}^{(g)}$]			& 100 & 1.9  & 50 &50 & $7\times 10^5$ & 15\,{\rm min}  \\
		CMB lensing [$B_{\ell_1\ell_2\ell_3}^{(\psi)}$]			&  100 &1.9 &50  & 50 & $7\times 10^5$ & 15\,{\rm min}  \\
		CMB anisotropies [$B_{\ell_1\ell_2\ell_3}^{(\Theta)}$]		& 70 & 0.1 &200 & 80 &  $10^5$& 10\,min \\
		\bottomrule
		\end{tabular}
	\end{center}
	\vspace{-0.2cm}
	\caption{Parameters used to compute the bispectra. $N_{\nu}$ is the number of frequencies used in the FFTlog, $b$ is the bias, $N_{\chi}$ and $N_r$ are the number of sampling points in the $\chi$ and $r$ integral, respectively and $N_{\ell_1\ell_2\ell_3}$ is the number of triangles computed. The last column corresponds to the time observed on a laptop using our {\sf Mathematica} code (without parallelization).}
	\label{Table:Bl}
\end{table}

\newpage
\addcontentsline{toc}{section}{References}
\bibliographystyle{utphys}
\bibliography{references}

\providecommand{\href}[2]{#2}\begingroup\raggedright\begin{thebibliography}{10}

\bibitem{LoVerde:2008re}
M.~LoVerde and N.~Afshordi, ``{Extended Limber Approximation},''
  \href{http://dx.doi.org/10.1103/PhysRevD.78.123506}{{\em Phys. Rev.}
  {\bfseries D78} (2008) 123506},
\href{http://arxiv.org/abs/0809.5112}{{\ttfamily arXiv:0809.5112 [astro-ph]}}.
%%CITATION = ARXIV:0809.5112;%%.

\bibitem{Campagne:2017xps}
J.~E. Campagne, J.~Neveu, and S.~Plaszczynski, ``{Angpow: A Software for the
  Fast Computation of Accurate Tomographic Power Spectra},''
\href{http://arxiv.org/abs/1701.03592}{{\ttfamily arXiv:1701.03592
  [astro-ph.CO]}}.
%%CITATION = ARXIV:1701.03592;%%.

\bibitem{Lesgourgues:2011re}
J.~Lesgourgues, ``{The Cosmic Linear Anisotropy Solving System (CLASS) I:
  Overview},''
\href{http://arxiv.org/abs/1104.2932}{{\ttfamily arXiv:1104.2932
  [astro-ph.IM]}}.
%%CITATION = ARXIV:1104.2932;%%.

\bibitem{DiDio:2013bqa}
E.~Di~Dio, F.~Montanari, J.~Lesgourgues, and R.~Durrer, ``{The CLASSgal Code
  for Relativistic Cosmological Large-Scale Structure},''
  \href{http://dx.doi.org/10.1088/1475-7516/2013/11/044}{{\em JCAP} {\bfseries
  1311} (2013) 044},
\href{http://arxiv.org/abs/1307.1459}{{\ttfamily arXiv:1307.1459
  [astro-ph.CO]}}.
%%CITATION = ARXIV:1307.1459;%%.

\bibitem{Liguori:2010hx}
M.~Liguori, E.~Sefusatti, J.~R. Fergusson, and E.~P.~S. Shellard, ``{Primordial
  non-Gaussianity and Bispectrum Measurements in the Cosmic Microwave
  Background and Large-Scale Structure},''
  \href{http://dx.doi.org/10.1155/2010/980523}{{\em Adv. Astron.} {\bfseries
  2010} (2010) 980523},
\href{http://arxiv.org/abs/1001.4707}{{\ttfamily arXiv:1001.4707
  [astro-ph.CO]}}.
%%CITATION = ARXIV:1001.4707;%%.

\bibitem{Hamilton:1999uv}
A.~J.~S. Hamilton, ``{Uncorrelated Modes of the Nonlinear Power Spectrum},''
  \href{http://dx.doi.org/10.1046/j.1365-8711.2000.03071.x}{{\em Mon. Not. Roy.
  Astron. Soc.} {\bfseries 312} (2000) 257--284},
\href{http://arxiv.org/abs/astro-ph/9905191}{{\ttfamily arXiv:astro-ph/9905191
  [astro-ph]}}.
%%CITATION = ASTRO-PH/9905191;%%.

\bibitem{ram2010}
W.~Research, ``Mathematica 8.0,'' 2010.

\bibitem{Regan:2010cn}
D.~M. Regan, E.~P.~S. Shellard, and J.~R. Fergusson, ``{General CMB and
  Primordial Trispectrum Estimation},''
  \href{http://dx.doi.org/10.1103/PhysRevD.82.023520}{{\em Phys. Rev.}
  {\bfseries D82} (2010) 023520},
\href{http://arxiv.org/abs/1004.2915}{{\ttfamily arXiv:1004.2915
  [astro-ph.CO]}}.
%%CITATION = ARXIV:1004.2915;%%.

\bibitem{Fergusson:2010ia}
J.~R. Fergusson, D.~M. Regan, and E.~P.~S. Shellard, ``{Rapid Separable
  Analysis of Higher Order Correlators in Large Scale Structure},''
  \href{http://dx.doi.org/10.1103/PhysRevD.86.063511}{{\em Phys. Rev.}
  {\bfseries D86} (2012) 063511},
\href{http://arxiv.org/abs/1008.1730}{{\ttfamily arXiv:1008.1730
  [astro-ph.CO]}}.
%%CITATION = ARXIV:1008.1730;%%.

\bibitem{Schmittfull:2016jsw}
M.~Schmittfull, Z.~Vlah, and P.~McDonald, ``{Fast Large-Scale Structure
  Perturbation Theory using One-Dimensional Fast Fourier Transforms},''
  \href{http://dx.doi.org/10.1103/PhysRevD.93.103528}{{\em Phys. Rev.}
  {\bfseries D93} no.~10, (2016) 103528},
\href{http://arxiv.org/abs/1603.04405}{{\ttfamily arXiv:1603.04405
  [astro-ph.CO]}}.
%%CITATION = ARXIV:1603.04405;%%.

\bibitem{McEwen:2016fjn}
J.~E. McEwen, X.~Fang, C.~M. Hirata, and J.~A. Blazek, ``{FAST-PT: A Novel
  Algorithm to Calculate Convolution Integrals in Cosmological Perturbation
  Theory},'' \href{http://dx.doi.org/10.1088/1475-7516/2016/09/015}{{\em JCAP}
  {\bfseries 1609} no.~09, (2016) 015},
\href{http://arxiv.org/abs/1603.04826}{{\ttfamily arXiv:1603.04826
  [astro-ph.CO]}}.
%%CITATION = ARXIV:1603.04826;%%.

\bibitem{Fergusson:2006pr}
J.~R. Fergusson and E.~P.~S. Shellard, ``{Primordial Non-Gaussianity and the
  CMB Bispectrum},'' \href{http://dx.doi.org/10.1103/PhysRevD.76.083523}{{\em
  Phys. Rev.} {\bfseries D76} (2007) 083523},
\href{http://arxiv.org/abs/astro-ph/0612713}{{\ttfamily arXiv:astro-ph/0612713
  [astro-ph]}}.
%%CITATION = ASTRO-PH/0612713;%%.

\bibitem{delaBella:2017qjy}
L.~F. de~la Bella, D.~Regan, D.~Seery, and S.~Hotchkiss, ``{The Matter Power
  Spectrum in Redshift Space using Effective Field Theory},''
\href{http://arxiv.org/abs/1704.05309}{{\ttfamily arXiv:1704.05309
  [astro-ph.CO]}}.
%%CITATION = ARXIV:1704.05309;%%.

\bibitem{Arfken:2005:MMP}
G.~B. Arfken and H.~J. Weber, {\em Mathematical Methods for Physicists}.
\newblock Elsevier, Oxford, 6th~ed., 2005.

\bibitem{Lemos:2017arq}
P.~Lemos, A.~Challinor, and G.~Efstathiou, ``{The Effect of Limber and Flat-Sky
  Approximations on Galaxy Weak Lensing},''
  \href{http://dx.doi.org/10.1088/1475-7516/2017/05/014}{{\em JCAP} {\bfseries
  1705} no.~05, (2017) 014},
\href{http://arxiv.org/abs/1704.01054}{{\ttfamily arXiv:1704.01054
  [astro-ph.CO]}}.
%%CITATION = ARXIV:1704.01054;%%.

\bibitem{Kitching:2016zkn}
T.~D. Kitching, J.~Alsing, A.~F. Heavens, R.~Jimenez, J.~D. McEwen, and
  L.~Verde, ``{The Limits of Cosmic Shear},''
\href{http://arxiv.org/abs/1611.04954}{{\ttfamily arXiv:1611.04954
  [astro-ph.CO]}}.
%%CITATION = ARXIV:1611.04954;%%.

\bibitem{Kilbinger:2017lvu}
M.~Kilbinger {\em et~al.}, ``{Precision Calculations of the Cosmic Shear Power
  Spectrum Projection},''
\href{http://arxiv.org/abs/1702.05301}{{\ttfamily arXiv:1702.05301
  [astro-ph.CO]}}.
%%CITATION = ARXIV:1702.05301;%%.

\bibitem{Kaiser:1987qv}
N.~Kaiser, ``{Clustering in Real Space and in Redshift Space},''
{\em Mon. Not. Roy. Astron. Soc.} {\bfseries 227} (1987) 1--27.
%%CITATION = MNRAA,227,1;%%.

\bibitem{Bonvin:2014xia}
C.~Bonvin, R.~Durrer, and R.~Maartens, ``{Can primordial magnetic fields be the
  origin of the BICEP2 data?},''
  \href{http://dx.doi.org/10.1103/PhysRevLett.112.191303}{{\em Phys. Rev.
  Lett.} {\bfseries 112} no.~19, (2014) 191303},
\href{http://arxiv.org/abs/1403.6768}{{\ttfamily arXiv:1403.6768
  [astro-ph.CO]}}.
%%CITATION = ARXIV:1403.6768;%%.

\bibitem{Dalal:2007cu}
N.~Dalal, O.~Dore, D.~Huterer, and A.~Shirokov, ``{The Imprints of Primordial
  non-Gaussianities on Large-Scale Structure: Scale-Dependent Bias and
  Abundance of Virialized Objects},''
  \href{http://dx.doi.org/10.1103/PhysRevD.77.123514}{{\em Phys. Rev.}
  {\bfseries D77} (2008) 123514},
\href{http://arxiv.org/abs/0710.4560}{{\ttfamily arXiv:0710.4560 [astro-ph]}}.
%%CITATION = ARXIV:0710.4560;%%.

\bibitem{Lewis:2006fu}
A.~Lewis and A.~Challinor, ``{Weak Gravitational Lensing of the CMB},''
  \href{http://dx.doi.org/10.1016/j.physrep.2006.03.002}{{\em Phys. Rept.}
  {\bfseries 429} (2006) 1--65},
\href{http://arxiv.org/abs/astro-ph/0601594}{{\ttfamily arXiv:astro-ph/0601594
  [astro-ph]}}.
%%CITATION = ASTRO-PH/0601594;%%.

\bibitem{Lewis:1999bs}
A.~Lewis, A.~Challinor, and A.~Lasenby, ``{Efficient Computation of CMB
  Anisotropies in Closed FRW Models},''
  \href{http://dx.doi.org/10.1086/309179}{{\em Astrophys. J.} {\bfseries 538}
  (2000) 473--476},
\href{http://arxiv.org/abs/astro-ph/9911177}{{\ttfamily arXiv:astro-ph/9911177
  [astro-ph]}}.
%%CITATION = ASTRO-PH/9911177;%%.

\bibitem{Seljak:1996is}
U.~Seljak and M.~Zaldarriaga, ``{A Line of Sight Integration Approach to Cosmic
  Microwave Background Anisotropies},''
  \href{http://dx.doi.org/10.1086/177793}{{\em Astrophys. J.} {\bfseries 469}
  (1996) 437--444},
\href{http://arxiv.org/abs/astro-ph/9603033}{{\ttfamily arXiv:astro-ph/9603033
  [astro-ph]}}.
%%CITATION = ASTRO-PH/9603033;%%.

\bibitem{DiDio:2016gpd}
E.~Di~Dio, H.~Perrier, R.~Durrer, G.~Marozzi, A.~M. Dizgah, J.~Nore{\~n}a, and
  A.~Riotto, ``{Non-Gaussianities due to Relativistic Corrections to the
  Observed Galaxy Bispectrum},''
  \href{http://dx.doi.org/10.1088/1475-7516/2017/03/006}{{\em JCAP} {\bfseries
  1703} no.~03, (2017) 006},
\href{http://arxiv.org/abs/1611.03720}{{\ttfamily arXiv:1611.03720
  [astro-ph.CO]}}.
%%CITATION = ARXIV:1611.03720;%%.

\bibitem{Desjacques:2016bnm}
V.~Desjacques, D.~Jeong, and F.~Schmidt, ``{Large-Scale Galaxy Bias},''
\href{http://arxiv.org/abs/1611.09787}{{\ttfamily arXiv:1611.09787
  [astro-ph.CO]}}.
%%CITATION = ARXIV:1611.09787;%%.

\bibitem{Fergusson:2009nv}
J.~R. Fergusson, M.~Liguori, and E.~P.~S. Shellard, ``{General CMB and
  Primordial Bispectrum Estimation I: Mode Expansion, Map-Making and Measures
  of $f_{\mathsmaller{\rm NL}}$},''
  \href{http://dx.doi.org/10.1103/PhysRevD.82.023502}{{\em Phys. Rev.}
  {\bfseries D82} (2010) 023502},
\href{http://arxiv.org/abs/0912.5516}{{\ttfamily arXiv:0912.5516
  [astro-ph.CO]}}.
%%CITATION = ARXIV:0912.5516;%%.

\bibitem{Fergusson:2008ra}
J.~R. Fergusson and E.~P.~S. Shellard, ``{The Shape of Primordial
  non-Gaussianity and the CMB Bispectrum},''
  \href{http://dx.doi.org/10.1103/PhysRevD.80.043510}{{\em Phys. Rev.}
  {\bfseries D80} (2009) 043510},
\href{http://arxiv.org/abs/0812.3413}{{\ttfamily arXiv:0812.3413 [astro-ph]}}.
%%CITATION = ARXIV:0812.3413;%%.

\bibitem{Bernardeau:2011tc}
F.~Bernardeau, C.~Bonvin, N.~Van~de Rijt, and F.~Vernizzi, ``{Cosmic Shear
  Bispectrum from Second-order Perturbations in General Relativity},''
  \href{http://dx.doi.org/10.1103/PhysRevD.86.023001}{{\em Phys. Rev.}
  {\bfseries D86} (2012) 023001},
\href{http://arxiv.org/abs/1112.4430}{{\ttfamily arXiv:1112.4430
  [astro-ph.CO]}}.
%%CITATION = ARXIV:1112.4430;%%.

\bibitem{book1}
W.~Magnus, F.~Oberhettinger, and R.~P. Soni, {\em Formulas and Theorems for the
  Special Functions of Mathematical Physics}.
\newblock Springer, Berlin, Heidelberg, 1966.

\bibitem{book2}
G.~Watson, {\em A Treatise on the Theory of Bessel Functions}.
\newblock CUP, 1966.

\bibitem{Press1996}
W.~H. Press, S.~a. Teukolsky, W.~T. Vetterling, and B.~P. Flannery, {\em
  {Numerical Recipes in Fortran 77: the Art of Scientific Computing. Second
  Edition}}, vol.~1.
\newblock 1996.

\end{thebibliography}\endgroup

\end{document}